\documentclass[fleqn,usenatbib]{mnras}

\usepackage{newtxtext,newtxmath}

\usepackage[T1]{fontenc}

\DeclareRobustCommand{\VAN}[3]{#2}
\let\VANthebibliography\thebibliography
\def\thebibliography{\DeclareRobustCommand{\VAN}[3]{##3}\VANthebibliography}

\usepackage{graphicx}	
\usepackage{amsmath}	
\usepackage{subfigure}

\usepackage{xspace}
\usepackage[table]{xcolor}

\newcommand{\lya}{Lyman-$\alpha$\xspace}
\newcommand{\lyaf}{Lyman-$\alpha$ forest\xspace}
\newcommand{\poned}{$P_{\rm 1D}$\xspace}
\newcommand{\ponedprime}{\ensuremath{P_{\rm 1D}^{\prime}}\xspace}

\newcommand{\lacehc}{\textsc{training}\xspace}
\newcommand{\simseed}{\textsc{seed}\xspace}
\newcommand{\simigm}{\textsc{reionisation}\xspace}
\newcommand{\simcurved}{\textsc{curved}\xspace}
\newcommand{\simh}{\textsc{growth}\xspace}
\newcommand{\simnu}{\textsc{neutrinos}\xspace}
\newcommand{\simns}{\textsc{running}\xspace}
\newcommand{\simcentral}{\textsc{central}\xspace}

\newcommand{\mflux}{\ensuremath{\bar{F}}\xspace}
\newcommand{\iMpc}{\ensuremath{\,\mathrm{Mpc}^{-1}}}

\defcitealias{Pedersen2021}{P21}

\newcommand{\lc}[1]{ {\color{black} #1}}

\title[A neural network emulator for the \lyaf 1D flux power spectrum]{A neural network emulator for the \lyaf 1D flux power spectrum}
\author[]{L.~Cabayol-Garcia$^{1,2}$\thanks{E-mail:lcabayol@pic.es},
J.~Chaves-Montero$^{1}$\thanks{E-mail: jchaves@ifae.es}, A.~Font-Ribera$^{1}$\thanks{E-mail: afont@ifae.es}, C.~Pedersen$^{3,4,5,6}$ 
\\ \\
$^{1}$Institut de F\'{\i}sica d'Altes Energies (IFAE), The Barcelona Institute of Science and Technology, 08193 Bellaterra (Barcelona), Spain \\
$^{2}$ Port d'Informaci\'{o} Cient\'{i}fica, Campus UAB, C. Albareda s/n, 08193 Bellaterra (Barcelona), Spain\\
$^{3}$ Courant Institute of Mathematical Sciences, New York University, New York, NY 10012, USA\\
$^{4}$ Centre for Data Science, New York University, New York, NY 10011, USA\\
$^{5}$ Center for Computational Astrophysics, New York, NY 10010, USA\\
$^{6}$ Department of Physics and Astronomy, University College London, London, WC1E 6BT, UK\\}

\date{Accepted XXX. Received YYY; in original form ZZZ}

\pubyear{2023}

\begin{document}
\label{firstpage}
\pagerange{\pageref{firstpage}--\pageref{lastpage}}
\maketitle

\begin{abstract}
The \lyaf offers a unique avenue for studying the distribution of matter in the high redshift universe and extracting precise constraints on the nature of dark matter, neutrino masses, and other $\Lambda$CDM extensions. However, interpreting this observable requires accurate modelling of the thermal and ionisation state of the intergalactic medium, and therefore resorting to computationally expensive hydrodynamical simulations. In this work, we build a neural network that serves as a surrogate model for rapid predictions of the one-dimensional \lya flux power spectrum (\poned), thereby making Bayesian inference feasible for this observable. Our emulation technique is based on modelling \poned as a function of the slope and amplitude of the linear matter power spectrum rather than as a function of cosmological parameters. We show that our emulator achieves sub-percent precision across the full range of scales ($k_{\parallel
}=0.1$ to $4\iMpc$) and redshifts ($z=2$ to 4.5) considered, and also for three $\Lambda$CDM extensions not included in the training set: massive neutrinos, running of the spectral index, and curvature. Furthermore, we show that it performs at the 1\% level for ionisation and thermal histories not present in the training set and performs at the percent level when emulating down to $k_{\parallel}=8\iMpc$. These results affirm the efficacy of our emulation strategy in providing accurate predictions even for cosmologies and reionisation histories that were not explicitly incorporated during the training phase, and we expect it to play a critical role in the cosmological analysis of the DESI survey.

\end{abstract}




\section{Introduction}

The \lyaf refers to a series of absorption features in the spectra of high-redshift quasars caused by the scatter of quasar light at 1216 \textup{\AA} of neutral hydrogen in the intergalactic medium \citep[IGM, for a review, see][]{mcquinn2016EvolutionIntergalacticMedium}. Consequently, the \lyaf is sensitive to density fluctuations as well as the thermal and ionisation state of the IGM, thereby containing precise cosmological and astrophysical information about redshifts well above the reach of large-scale galaxy surveys.

Cosmological analyses of the \lyaf rely on either three-dimensional correlations of the \lya transmission field to measure baryonic acoustic oscillations \citep[BAO, e.g.,][]{busca2013LyaBAODR9,
slosar2013MeasurementBaryonAcoustica, 
delubac2015LyaBAODR11,
bautista2017MeasurementBaryonAcoustic,
desainteagathe2019BaryonAcousticOscillations,
dumasdesbourboux2020CompletedSDSSIVExtended}, or correlations along the line-of-sight of each individual quasar \citep[one-dimensional flux power spectrum, \poned, e.g.,][]{mcdonald2006LyaForestPowera, p1d_PalanqueDelabrouille2013, p1d_Chabanier2019} to set constraints on the sum of neutrino masses \citep{palanque-delabrouille2015ConstraintNeutrinoMasses, emuparamv2_PD2, yeche2017ConstraintsNeutrinoMasses, palanque-delabrouille2020HintsNeutrinoBounds}, the nature of dark matter \citep{baur2016LymanalphaForestsCoola, baur2017ConstraintsLyaForests, yeche2017ConstraintsNeutrinoMasses, armengaud2017ConstrainingMassLight, irsic2017FirstConstraintsFuzzy, palanque-delabrouille2020HintsNeutrinoBounds,gpemucosmo_rogers}, and even the reionisation \citep{zaldarriaga2001ConstraintsLyaForest, meiksin2009PhysicsIntergalacticMedium, lee2015IGMConstraintsSDSSIII, mcquinn2016EvolutionIntergalacticMedium} and thermal history \citep{viel2006CosmologicalAstrophysicalParameters, bolton2008PossibleEvidenceInverted} of the Universe.

Over the last decade, first the Baryon Oscillation Spectroscopic Survey \citep[BOSS;][]{boss_dawson2013} and then the Extended Baryon Oscillation Spectroscopic Survey \citep[eBOSS;][]{eboss_dawson2016} have dramatically increased the number of \lyaf measurements available, thereby significantly increasing the precision of cosmological and IGM constraints \citep{dumasdesbourboux2020CompletedSDSSIVExtended}. 
The ongoing Dark Energy Spectroscopic Instrument survey \citep[DESI,][]{DESI_collab2016} will quadruple the number of line-of-sights, which will increase even more the constraining power of \lyaf analyses \citep{font-ribera2014DESIOtherDark}. 

BAO measurements of the \lyaf only use the correlations on large scales that can be modelled with linear theory. On the other hand, extracting constraints from the \poned measurements is quite challenging because they are sensitive to the complex physical processes that affect the distribution of neutral hydrogen in the IGM, and thus a precise interpretation of this observable requires resorting to time-consuming hydrodynamical simulations \citep{hydro_Cen1994, hydro_Miralda1996, hydro_Meiksin2001, hydro_Lukic2015, hydro_Walther2021, hydro_Chabanier2023}.

Bayesian inference techniques require of the order of $10^6$ evaluations to set robust constraints on cosmological parameters; consequently, a traditional analysis would require running the same number of hydrodynamical simulations, each of these taking more than $\sim 10^5$ CPU hours, making Bayesian inference unfeasible. One solution to this problem is constructing fast surrogate models that interpolate predictions to regions of the parameter space not sampled by simulations \citep[for the first application of this technique in cosmology, see][]{heitmann2006CosmicCalibrationa, habib2007CosmicCalibrationConstraints}. As a result, the number of simulations required for Bayesian inference decreases dramatically from millions to dozens or hundreds. We will refer to these models as emulators hereafter.

The two main approaches to building emulators are to use architectures based on a Gaussian process \citep[GP;][]{sacks1989DesignAnalysisComputer, mackay1998introduction} or a neural network \citep[NN;][]{mcculloch1943logical}. These two techniques are different types of supervised learning algorithms, with the first and second typically considered as ``non-parametric'' and ``over-parameterized'', respectively. In general, GPs require less training data than NNs and produce more robust predictions, but this comes at the cost of presenting runtimes that scale with the cube of the number of data points instead of linearly like NNs. As a result, GPs and NNs are more appropriate for small and large datasets, respectively. These two architectures have been used for multiple applications, e.g., GPs for emulating the matter power spectrum \citep[e.g.,][]{heitmann2009CoyoteUniverseII, lawrence2010CoyoteUniverseIII, heitmann2016MiraTitanUniversePrecision, lawrence2017MiraTitanUniverseII}, the halo mass function \citep[e.g.,][]{bocquet2020MiraTitanUniverseIII}, and the one-dimensional \lya flux power spectrum \citep[e.g.,][]{emugp_bird2019, emugp_rogers2019, emugp_Walther2019, Pedersen2021, emugp_Takhtaganov2021, emugp_rogers2021, gpemu:P1DFernandez2022, bird2023PRIYANewSuite} or NNs for also emulating the matter power spectrum \citep[e.g.,][]{emu_angulo2021, emu_arico2021}, galaxy clustering and galaxy-galaxy lensing \citep[e.g.,][]{darkemulator, chaves-montero2023GalaxyFormationOrigin}, \lc{one-dimensional power spectrum of the \lyaf} \citep{ennemu:P1DMolaro2023}, and accelerating the predictions of Einstein-Boltzmann Solvers \citep{emunn_gunther, emunn_Nygaard}.

In this paper, we build the first NN-based emulator for the one-dimensional power spectrum of the \lyaf \lc{as a function of cosmology and the IGM state}. We create it following the approach devised by \citet[][\citetalias{Pedersen2021} hereafter]{Pedersen2021}, which relies on emulating \poned as a function of the amplitude and slope of the linear matter power spectrum on small scales rather than as a function of cosmological parameters. The main advantage of this approach is that it reduces the dimensionality of the problem, and it enables precise predictions for redshifts, cosmological parameters, and $\Lambda$CDM extensions not considered in the training set \citep[][]{Pedersen2021, pedersen2023CompressingCosmologicalInformation}.
Our main motivation for building an NN- instead of a GP-based emulator like \citetalias{Pedersen2021} is that they can handle larger training datasets, like the one that will be needed to accurately interpret \poned measurements from DESI. The emulator developed in this paper is publicly accessible at \url{https://github.com/igmhub/LaCE}.

The outline of this paper is as follows. Section \ref{sec:methods} presents the hydrodynamical simulations, their post-processing, and the \poned parametrisation. In \S\ref{sec:emulator} and \S\ref{sec:characterisation}, we present the neural-network emulator developed in this paper and characterise its training procedure and training sample. Section \ref{sec:results} contains the main results obtained with the emulator. This includes testing the \lacehc simulations, validating the emulator at arbitrary redshifts, and testing it on simulations with different cosmology and astrophysics.
Section \ref{sec:extended_emulator} presents an extended version of the emulator to smaller scales. Finally, \S\ref{sec:conclusions} concludes the paper.

\section{Methods}
\label{sec:methods}

In this section, we describe the simulations from which we extract \poned measurements for calibrating and testing our emulator in \S\ref{sec:methods_sims}, how we extract these measurements in \S\ref{sec:methods_post}, and the input parameters for our emulator in \S\ref{sec:methods_params}.


\begin{figure}
\includegraphics[width=\columnwidth]{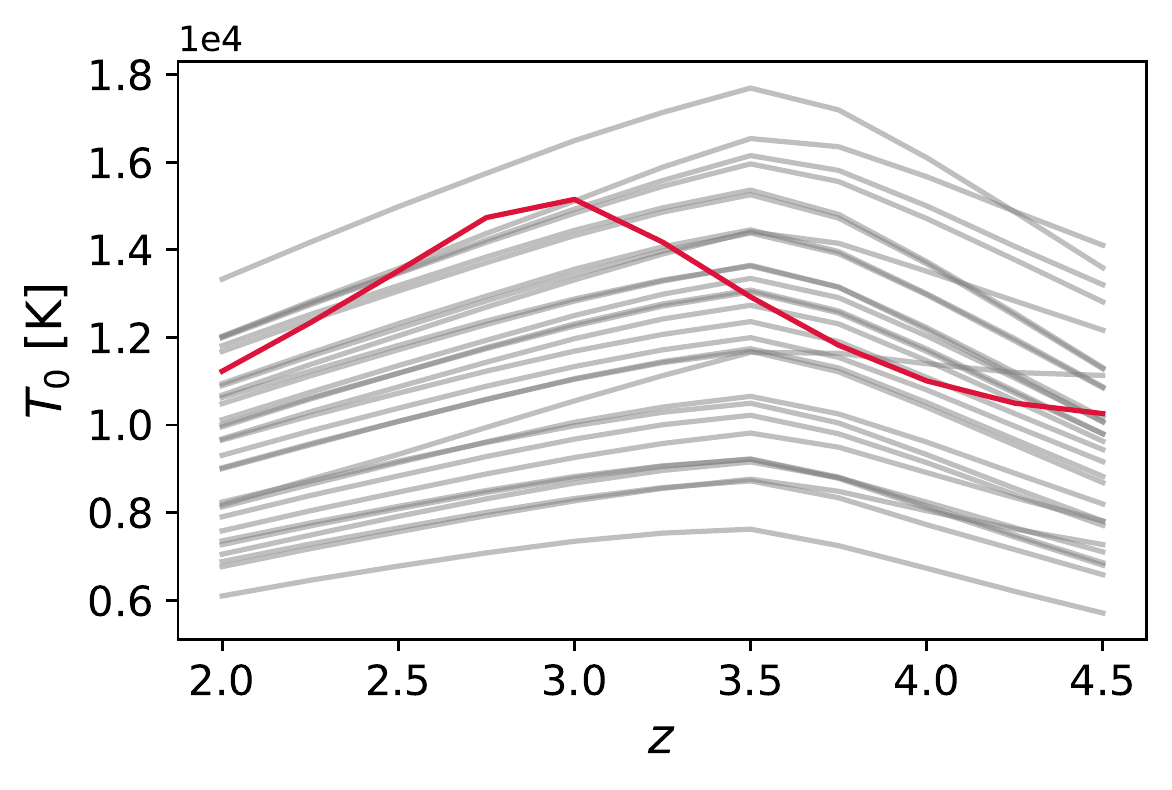}
\centering
\caption{Redshift evolution of the gas temperature at mean density. The gray lines show the results for \lacehc simulations, while the red line does so for the \simigm simulation. The thermal histories of the \lacehc simulations, which are used to train the \poned emulator, are significantly different from that of the \simigm simulation, which is used for testing.}
\label{fig:T0_vs_z}
\end{figure} 

\subsection{Simulations}
\label{sec:methods_sims}

We train our emulator using \poned measurements from a suite of 60 flat $\Lambda$CDM cosmological hydrodynamical simulations described in detail in \citetalias{Pedersen2021}; throughout this work, we refer to these simulations as \lacehc. The simulations were run employing \textsc{mp-gadget}\footnote{\url{https://github.com/MP-Gadget/MP-Gadget/}} \citep{feng2018MpGadgetMpGadgetTag, emugp_bird2019}, a massively scalable version of the cosmological structure formation code \textsc{gadget-3} \citep[last described in][]{Gadget_Springel}. Each simulation tracks the evolution of $768^3$ dark matter and baryon particles from $z=99$ to $z=2$ inside a simulation box of $L = 67.5$ Mpc on a side and generates 11 output snapshots uniformly spaced in redshift between $z=4.5$ and $z=2$. 
\lc{The selected values strike a balance between resolving a sufficiently large range of scales for analysing DESI data \citep{Ravoux2023,karaçayli2023} and limitations of computational resources.}

To increase computational efficiency, star formation is included using a simplified prescription that turns regions of baryon overdensity $\Delta_\mathrm{b}>1000$ and temperature $T<10^5$ K into collisionless stars, which is justified by the negligible contribution of high-density regions to the Ly$\alpha$ forest \citep[e.g.,][]{viel2004ConstraintsPrimordialPower}. Also for efficiency purposes, the simulations use a spatially uniform ultraviolet background implementation from \citet{haardt2012RadiativeTransferClumpya} and do not include active galactic nuclei (AGN) feedback; the first and second approximations may lead to up to $\sim10\%$ errors on \poned predictions especially at high redshift \citep[$z\simeq4$, e.g.,][]{pontzen2014ScaledependentBiasBaryonicacousticoscillationscale,gontchoagontcho2014EffectIonizingBackground, suarez2017LargescaleFluctuationsCosmic} and low redshift \citep[$z\simeq2$, e.g.,][]{chabanier2020ImpactAGNFeedback}, respectively. 
\lc{These approximations are justified because we focus on emulating the ``\lya only'' power spectrum throughout this work, without attempting to model astrophysical contaminants like AGN feedback, damped Lyman-alpha absorbers (DLAs), or metal absorbers. 
These contaminants are often modelled with simple prescriptions, and are added to the ``\lya only'' prediction before comparing to the \poned measurement 
\citep{p1dparams_McDonald,palanque-delabrouille2015ConstraintNeutrinoMasses,palanque-delabrouille2020HintsNeutrinoBounds}.
\footnote{One exception is the recent work by \cite{bird2023PRIYANewSuite}, that incorporates some of the astrophysical contaminants already at the emulator level.}}

The \lacehc simulations adopt 30 different sets of cosmological and astrophysical parameters selected according to a Latin hypercube design so the space of interest is sampled efficiently \citep{mckay1979ComparisonThreeMethods}. Two realisations were run for each combination using an initial mode amplitude fixed to the ensemble mean and opposite Fourier phases \citep{angulo2016CosmologicalNbodySimulations, pontzen2016InvertedInitialConditions}. These initial conditions, commonly known as ``fixed-and-paired'', significantly reduce cosmic variance in the Ly$\alpha$ forest power spectrum \citep{fixedpaired_Villaescusa, anderson2019CosmologicalHydrodynamicSimulations}; throughout the remainder of this work, we refer to measurements from the simulations of a pair as from different phases. The impact of cosmic variance on \lacehc simulations is nevertheless considerable due to their limited size. To separate this source of uncertainty from others, all simulation pairs were run using the same distribution of Fourier phases.

Motivated by the emulation strategy, the simulations explore different values of the amplitude and slope of the linear power spectrum,
\begin{align}
    \label{eq:amplitude}
    & \Delta_\mathrm{p}^2(z) = k^3 P_{\rm lin} (k_\mathrm{p}, z),\\
    \label{eq:slope}
    & n_\mathrm{p}(z) = \left(\mathrm{d}\log P_{\rm lin} / \mathrm{d}\log k\right)\mid_{k = k_\mathrm{p}},
\end{align}
where $k_\mathrm{p}$ is the pivot scale at which these are computed and $P_{\rm lin}$ is the linear power spectrum of cold dark matter and baryons\footnote{Note that $P_{\rm lin}$ does not include the contribution of neutrinos for cosmologies with massive neutrinos.}. Specifically, the simulations use values within the ranges $\Delta^2_\mathrm{p}(z=z_\star) \in [0.25,\, 0.45]$ and $n_\mathrm{p}(z=z_\star) \in [-2.35,\, -2.25]$, which are defined at $z_\star=3$ and $k_\mathrm{p}=0.7\,\mathrm{Mpc}^{-1}$ because this redshift and scale are approximately at the centre of the ranges of interest for DESI \lya studies (Ravoux et al. in prep, Kara{\c{c}}ayl{\i} et al. in prep.). As for the other cosmological parameters, the simulations use the same value for the Hubble parameter ($H_0=67\,\mathrm{km}\,\mathrm{s}^{-1} \mathrm{Mpc}^{-1}$), physical cold dark matter density ($\omega_\mathrm{c} \equiv \Omega_\mathrm{c} h^2=0.12$), and physical baryon density ($\omega_\mathrm{b} \equiv \Omega_\mathrm{b} h^2=0.022$), where $h=0.67$ is the dimensionless Hubble parameter. Note that \citetalias{Pedersen2021} and \citet{pedersen2023CompressingCosmologicalInformation} showed that our emulation strategy produces precise results for simulations with cosmological parameters outside of the training set and $\Lambda$CDM extensions; we test this further in \S\ref{sec:other_cosmo}.

The \lacehc simulations consider three astrophysical parameters to account for uncertainties in the reionisation and thermal history of the Universe. We consider as a fiducial model the histories from \citet{haardt2012RadiativeTransferClumpya}, and we then vary the value of the previous 3 parameters to perturb these histories following prescriptions from \citet{onorbe2017SelfconsistentModelingReionization}. Specifically, the simulations explore $z_\mathrm{H}\in[5.5,\,15]$, which indicates the midpoint of hydrogen reionisation, while holding fixed the redshift of the second helium ionisation to $z_{\ion{He}{II}}=3.5$. In addition, these use $H_\mathrm{A}\in[0.5,\,1.5]$ and $H_\mathrm{S}\in[0.5,\,1.5]$, which account for the uncertain effect of helium reionisation on the IGM temperature by rescaling the \ion{He}{II} photo-heating rate, $\epsilon_0$, such that $\epsilon=H_\mathrm{A} \Delta_\mathrm{b}^{H_\mathrm{S}}\epsilon_0$. In this way, the thermal state of the IGM is correctly coupled to the gas pressure.

In addition, we use seven pairs of simulations with cosmological and astrophysical parameters not considered in the \lacehc simulations for evaluating different aspects of the emulation strategy\footnote{Some of these simulations were already described in \citetalias{Pedersen2021} and \citet{pedersen2023CompressingCosmologicalInformation} but named differently. The simulations \simh and \simnu were referred to as $h$ and $\nu$ in \citetalias{Pedersen2021} and the \simseed simulation as \textit{diff seed} in \citet{pedersen2023CompressingCosmologicalInformation}.}. The first is the \simcentral simulation, which uses parameters at the centre of the \lacehc parameter space and serves to evaluate the performance of the emulator in optimal conditions. The second is the \simseed simulation, with the same parameters as the \simcentral simulation but different initial conditions, which serves to characterise the impact of cosmic variance on the results.

We also use the \simh, \simnu, \simns, and \simcurved simulation pairs, which present the same amplitude and slope of the matter power spectrum at $z=3$, physical CDM and baryonic densities, and astrophysical parameters as the \simcentral simulation, but the \simh simulation uses 9\% larger Hubble constant and 18\% smaller $\Omega_\mathrm{M}$, the \simnu simulation includes massive neutrinos ($\sum m_\nu=0.3$ eV) implemented using the linear response approximation from \citet{ali-haimoud2013EfficientImplementationMassive}, the \simns simulation uses a non-zero running of the primordial power spectrum slope ($\mathrm{d}\,n_s/\mathrm{d}\log k=0.015$), and the \simcurved simulation considers an open universe ($\Omega_k=0.03$). These simulations serve to test the precision of the emulation strategy for cosmologies not included in the training set.

Furthermore, we consider the \simigm simulation, with the same cosmological parameters as the \simcentral simulation but implementing an ionisation history from \citet{puchwein2019ConsistentModellingMetagalactic}. The main difference between this and the \lacehc simulations is that the helium ionisation history of the first peaks at a later time, which translates into different IGM thermal histories. In Fig.~\ref{fig:T0_vs_z}, the grey lines display the thermal histories of all \lacehc simulations, while the red line does so for the \simigm simulation. As expected, we can see that the thermal history of the \simigm simulation peaks at a later time relative to those of the \lacehc simulations due to the also lower $z_{\ion{He}{II}}$ for the first. This simulation serves to test the performance of the \poned emulator for ionisation and thermal histories different from those used in the training set.


\subsection{Post-processing}
\label{sec:methods_post}

We extract \poned measurements from the simulations described in \S\ref{sec:methods_sims} as follows. For each simulation, we first consider one of the simulation axes as the line of sight, and then we displace particles from real to redshift space along this axis. We continue by computing the transmitted flux fraction along $768^2$ uniformly-distributed line of sights along this axis using \textsc{FSFE}\footnote{\url{https://github.com/sbird/fake_spectra}} \citep{bird2017FSFEFakeSpectra}; these lines of sights are commonly known as skewers. The line-of-sight resolution of the skewers is set to 0.05 Mpc, which is enough to resolve the thermal broadening scale. Then, we compute the Fourier transform of the transmitted flux fraction for each skewer, and we estimate \poned by taking the average of the Fourier transform of all skewers. Finally, we iterate over the two remainder simulation axes. By doing so, we sample different directions of the velocity field, extracting further information from the simulations.

We repeat the previous procedure for each simulation and snapshot, ending up with 30 (cosmologies) $\times$ 2 (opposite Fourier phases) $\times$ 3 (simulation axes) $\times$ 11 (snapshots) $=1980$ \poned measurements. Additionally, in post-processing, we vary the mean flux of the snapshots by scaling the effective optical depth of the skewers to 0.90, 0.95, 1.05, and 1.10 times its original value \citep[see][for more details about this approach]{hydro_Lukic2015}, and then we recompute the power spectrum, ending up with 9900 \poned measurements in total. As a result, this post-processing presents a few improvements compared to that carried out by \citetalias{Pedersen2021}: three simulation axes instead of one, $768^2$ skewers instead of $500^2$, and mean flux rescalings.


\subsection{Emulator parameterisation}
\label{sec:methods_params}

We use the amplitude and slope of the linear matter power spectrum at $k_\mathrm{p}=0.7\,\mathrm{Mpc}^{-1}$ to capture the cosmological dependence of \poned measurements. This is justified because, when rescaling the linear matter power spectra of cosmologies within \textit{Planck} priors so these match the amplitude and slope of the best-fitting \textit{Planck} solution, the amplitude of the variations is smaller than 1\% \citepalias[see fig.~1 of][]{Pedersen2021}. It is also important to note that the \lyaf probes cosmic times during which the universe is practically Einstein de-Sitter, and for such universe, the growth rate of velocities (and therefore redshift space distortions) has no cosmological dependence.

We use four parameters to describe the astrophysical dependence of the measurements. The first is the mean transmitted flux fraction \mflux, which encodes information about the ionisation state of the gas and is related to the effective optical depth as $\tau_{\rm eff} = -\log \mflux$. The next two inform us about the thermal state of the gas probed by the \lyaf. The first, $\gamma$, describes the slope of the temperature-density relation \citep[e.g.,][]{hydro_Lukic2015}, $T = T_0 \Delta_\mathrm{b}^{\gamma -1}$, where $T_0$ is the gas temperature at mean density. The second, $\sigma_\mathrm{T}$, captures the thermal broadening of absorption features due to the thermal motion of gas,
\begin{equation}
    \sigma_\mathrm{T} = \sigma_\mathrm{T,0} \sqrt{\frac{T_0 [K]}{10^4}}\frac{1+z}{H(z)}\,,
\end{equation}
where $\sigma_\mathrm{T,0}=9.1$ km s$^{-1}$. Finally, the fourth parameter, $k_\mathrm{F}$, captures the pressure smoothing scale \citep{kulkarni2015CharacterizingPressureSmoothing}. Note that we use $\sigma_\mathrm{T}$ and $k_\mathrm{F}$ in inverse comoving units \citepalias[for more details, see][]{Pedersen2021}.


\section{Emulators}
\label{sec:emulator}

In this section, we describe the two \poned emulators that we use throughout the paper: a GP-based emulator (\texttt{LaCE-GP}) in \S\ref{sec:emulator:gp} and an NN-based emulator (\texttt{LaCE-NN}) in \S\ref{sec:emulator:nn}. We provide a brief description of the first, which was already discussed in \citetalias{Pedersen2021}, and we focus on the second, which is the main contribution of this work.

\subsection{Gaussian process emulator (\texttt{LaCE-GP})}
\label{sec:emulator:gp} 

The \texttt{LaCE-GP} emulator maps the parameter space $\Vec{\theta} = [\Delta_\mathrm{p}^2, n_\mathrm{p}, \mflux, \sigma_\mathrm{T}, \gamma, k_\mathrm{F}$] to the \poned measurements (see \S\ref{sec:methods_params}) based on a Python-based implementation using the package \textsc{Gpy}. Specifically, it normalises input parameters $\Vec{\theta}$ so these vary within a unit volume and \poned by the median of all \poned used. The original version of this emulator \citepalias{Pedersen2021} predicts the value of \poned for each input scale $k_{\parallel
}$. The emulator was later updated to predict the best-fitting coefficients of a fourth-degree polynomial describing \poned measurements because \citet{pedersen2023CompressingCosmologicalInformation} found that the original implementation was significantly affected by cosmic variance.

In this work, we used a modified version of the \citet{pedersen2023CompressingCosmologicalInformation} emulator: we train it using the new post-processing of the LaCE simulations (\S\ref{sec:methods_post}), extend the fitting range from $k_{\parallel
}=3$ to $k_{\parallel}=4\ {\rm Mpc}^{-1}$, and predict the coefficients of a fifth-order polynomial (to account for the extended range of scales, see Appendix~\ref{sec:poly_order}). In \S\ref{sec:other_comparison}, we provide a detail comparison between the performance of this and the \texttt{LaCE-NN} emulator.

 \subsection{Neural network emulator (\texttt{LaCE-NN})}
\label{sec:emulator:nn}

\begin{figure*}
\includegraphics[width= 0.98\textwidth]{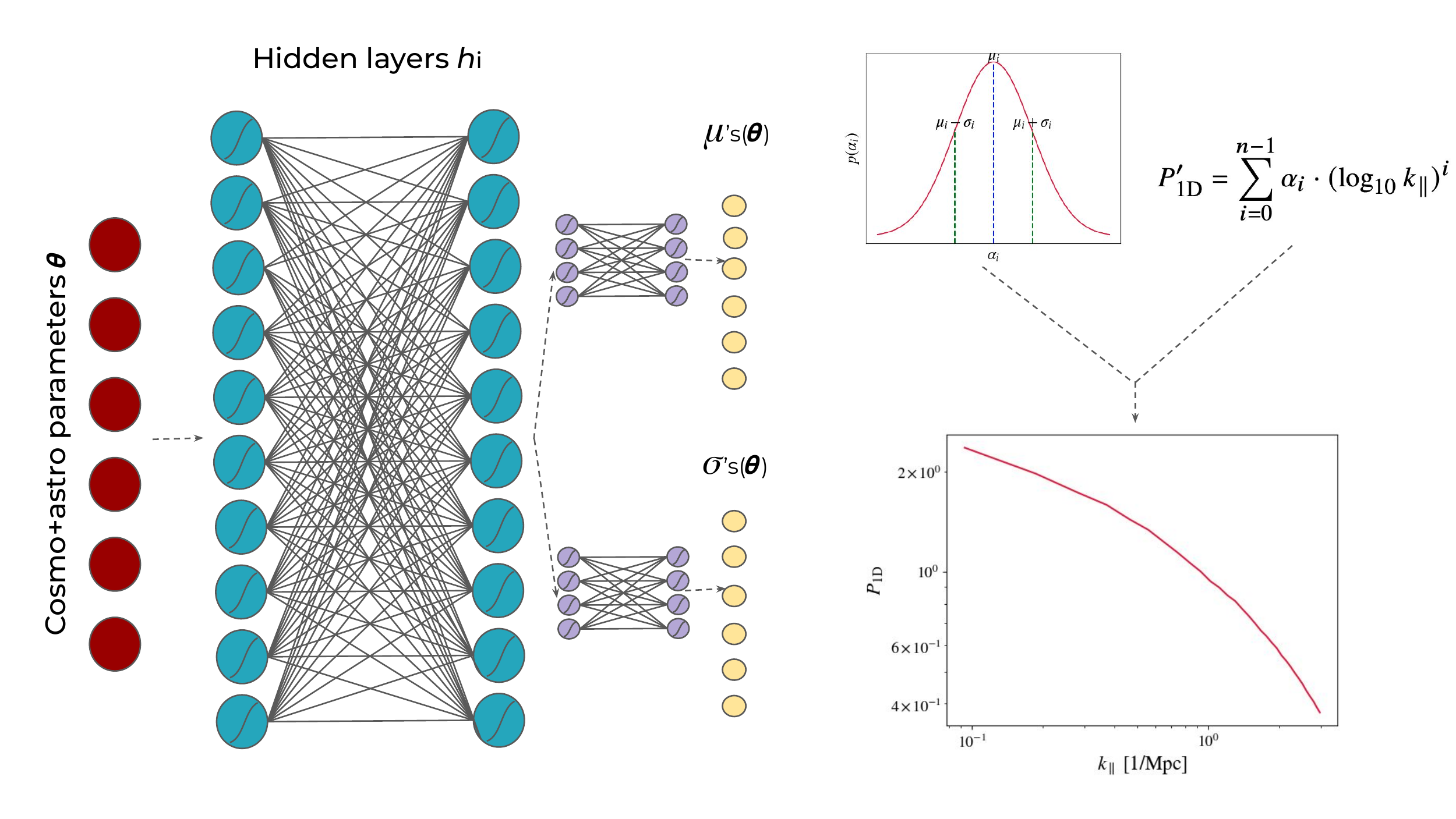}
\centering
\caption{Architecture of the mixture density network used in this work. The red circles correspond to the six input parameters (\S\ref{sec:methods_sims}), and the blue circles and their connections represent a set of shared hidden layers. The symbol within the blue circles represent the activation function, which in our case is a Leaky-ReLU. The output of these hidden layers is the input of two independent sets of hidden layers, each of them mapping to six output parameters. These output parameters parametrise the mean and standard deviation of six Gaussian distributions corresponding to the six polynomial coefficients of the \poned fit (Eq.\ref{eq:p1d_polynomial}.)}
\label{fig:net_architecture}
\end{figure*}

\begin{figure}
\includegraphics[width= 0.48\textwidth]{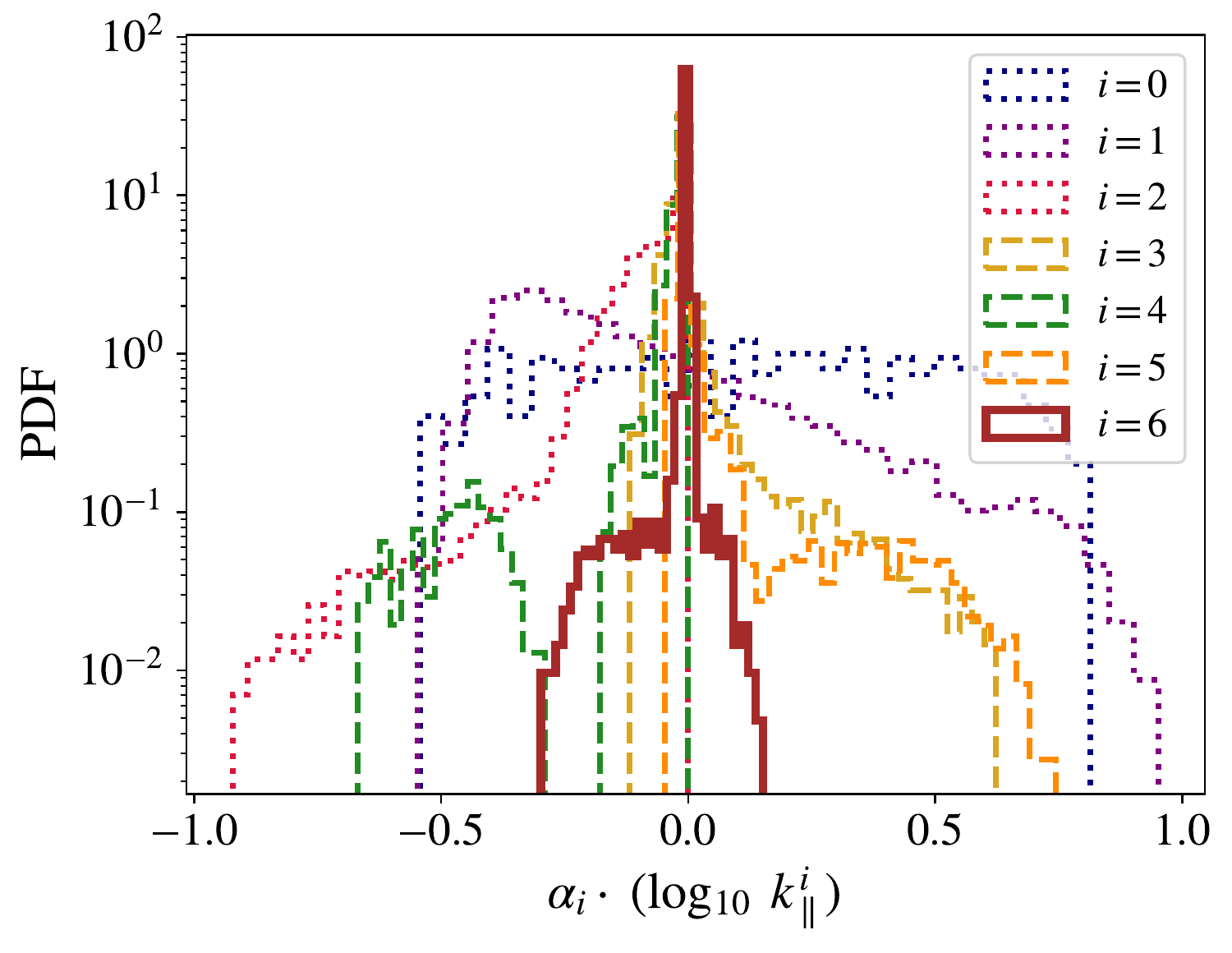}
\centering
\caption{Contribution of each polynomial term to the \poned reconstruction. Each colour corresponds to the histogram of the contribution of each polynomial order computed for all snapshots in the \lacehc simulations (\S\ref{sec:methods_sims}). The zero- and first-order coefficients ($i=0$ and $i=1$, blue and purple dotted lines) gently contribute to the \poned. The second- and third-order coefficients compensate for each other ($j=2$ and $j=3$, red and yellow lines). Similarly, the fourth and fifth-order coefficients ($i=4$, $i$ =5, green and orange dashed lines) also compensate for each other. Finally, the sixth-order contribution ($i=6$, solid brown) is centred at zero.}
\label{fig:coeff_distribution}
\end{figure}

Neural networks are mathematical models composed of interconnected layers of nodes, each of which possesses trainable weights and biases ($w$, $b$). The initial and final layers are commonly referred to as the input and output layers, respectively, and they have the same number of nodes as the dimensions of the input and output data. The intermediate layers are known as hidden layers and their number of nodes is a flexible hyperparameter of the network. Increasing the number of layers enhances the network's complexity, enabling it to handle more challenging tasks. However, networks with a higher number of layers are more susceptible to overfitting, which occurs when an excessively intricate model captures irrelevant patterns in the training data, leading to impaired generalization on new data.

In practice, the input data to the layer $i$ ($\vec{x_i}$) is multiplied by the set of trainable weights associated with the nodes in such layer, and the resulting values are passed through a non-linear function
\begin{equation}
    \vec{y_i} = g( \vec{x_i} \cdot \vec{w_i} + \vec{b_i}),
    \label{eq:forward_prop}
\end{equation} 
where $g(\cdot)$ is commonly known as the activation function and $\vec{y_i}$ represents the output of this layer. The training process involves finding the optimal weights and biases that minimise the difference between the network's output and the target "true" values. This discrepancy is quantified using a loss function ($\mathcal{L}$) and the optimisation is achieved using an optimisation algorithm, such as gradient descent. This algorithm calculates the gradient of the loss function with respect to the weights and biases and adjusts them in a manner that minimises the loss. 

In this paper, we  use a mixture-density network \citep[MDN,][]{MDN_Bishop1994}, a type of neural network that can model complex probability distributions. MDNs use a combination of neural network architecture and mixture models to estimate the parameters of a probability distribution from the input data. Each mixture component is associated with the mean and variance of a (typically) Gaussian distribution, and the network optimises such parameters to best describe the probability distribution of the target prediction.

Fig.~\ref{fig:net_architecture} shows the MDN used in this work, which is written in \textsc{PyTorch}. It maps the six parameters $\Vec{\theta} = [\Delta_p^2, n_p, \mflux, \sigma_T, \gamma, k_F]$ described in \S\ref{sec:methods_params} (red circles) to the probability distribution of six polynomial coefficients best fitting the \poned, where we assume the polynomial coefficients to follow a Gaussian around the true value. 
\lc{By employing an MDN, we gain the capability to effectively capture and represent complex distributions of polynomial coefficients. However, the emulator reaches sub-percent accuracy and reasonable uncertainties when assuming Gaussian polynomial coefficients. We consider this approximation to be a favourable choice as it eases the training process while yielding satisfactory results. In App.\ref{sec:covariance}, we show the covariance of the emulated \poned and compare it with the error covariance, which corresponds to the covariance of $P_{\rm 1D}^{\rm Pred}/P_{\rm 1D}^{\rm True}$. Although further investigations are needed, we attribute the main source of discrepancy between the two covariances to the assumption that polynomial coefficients are independent rather than the Gaussianity assumption. }

The input layer of the emulator maps the six input parameters to ten hidden variables (6:10) and it is then followed by a hidden space with four hidden layers mapping ten input to 100 output parameters (blue circles). The 100 output parameters of the hidden layers are the input of two independent sets of layers with architecture 100:50:5 (soft-purple circles) mapping the output of the previous hidden space to the mean and standard deviation of the Gaussian distribution best describing the six polynomial coefficients (yellow circles). 

\lc{The MDN is trained with an \texttt{Adam} optimiser \citep{adam_Diederik2015} for 100 epochs with an initial learning rate of $10^{-3}$ that decreases by a factor of ten after 75 epochs. To optimise the performance of the neural network, we conducted a series of fine-tuning steps using the hypercube training simulations and utilising the central simulation as a validation set. Regarding the network architecture, we initially employed a shallow network with a minimal number of parameters. Gradually, we increased the depth by adding more layers and parameters, evaluating the performance improvement achieved at each step. We continued this process until we reached a point where adding more layers and parameters did not yield substantial gains. The architecture tests were carried out with the other hyperparameters fixed to default values already reaching low residual errors on the central simulation. }

\lc{For the other hyperparameters, we first experimented with different learning rates, specifically exploring values of $10^{-2}$, $10^{-3}$, and $10^{-4}$. The loss function with a learning rate of $10^{-2}$ is unstable and does not converge while for $10^{-4}$ the convergence is very slow. Next, we investigated the relationship between the number of epochs and the residual error achieved on the validation set. By analysing how the residual error curve evolves with increasing epochs, we determined the optimal number of epochs at which the curve converges. This task proved to be more challenging due to its high correlation with the learning-rate scheduler. This is a technique to adjust the learning rate over time during the training to improve the performance of the model. Hence, we performed a grid search, systematically evaluating various combinations of epochs and the scheduler parameter. We finally identified 100 epochs with a learning-rate scheduler reducing the learning rate by a factor of 10 after 75 epochs as the most optimal combination. We repeated the training with these parameters several times to make sure that the emulator has always converged after such a number of iterations. Additionally, we explored different optimisers in an attempt to improve the network's performance on the central simulation. Although we tested several optimisers, (e.g. \texttt{SGD}, \texttt{Adam}, \texttt{AdamW}) we did not observe significant differences in their impact.}

The neural network fits the polynomial form
\begin{equation}
    P_{\rm 1D}^{\prime} = \sum_{i=0}^{n-1} \alpha_i \cdot (\log_{10}{k_{\parallel}})^i\, ,
    \label{eq:p1d_polynomial}
\end{equation} where $n$ is the order of the polynomial and $P_{\rm 1D}^{\prime}$ is the $\log_{10}$\,\poned scaled by the median of all \poned measurements in the training sample ($A_{\rm scalings}$), i.e.
\begin{equation}
    P_{\rm 1D}^{\prime} = \log_{\rm 10}{\left(P_{\rm 1D} / A_{\rm scalings}\right)}\,.
    \label{eq:p1d_scaling}
\end{equation}  The network also predicts an uncertainty for each polynomial coefficient ($\sigma_i$ ) that propagates to an uncertainty for $\log_{10} P_{\rm 1D}^{\prime}$
\begin{equation}
    \sigma_{\log P_{\rm 1D}}^2 = \sum_{i=0}^{n-1} \sigma_i^2 \cdot (\log_{10}{k_{\parallel}})^{2i}\, .
    \label{eq:p1derr_polynomial}
\end{equation} In App.~\ref{sec:covariance}, we also explore MonteCarlo methods to obtain a covariance matrix from the distribution of the polynomial coefficients. 

To fit the \poned in the $k_{\parallel}$ range $(0,4]\ \rm{Mpc}^{-1}$, we use a fifth-order polynomial  ($n=5$), which is justified in App.\,\ref{sec:poly_order}. To further back up this decision,  Fig.\ref{fig:coeff_distribution} shows the contribution of the optimal polynomial terms once they have been optimised by the emulator up to sixth degree. The zero- and first-order coefficients broadly contribute to the \poned. Then, second- and third-order contributions compensate for each other and a similar thing happens with order four and five. Finally, the sixth order term is already centred at zero, and therefore its contribution to the \poned measurements is smaller.

After each training iteration, we compare the emulator predictions (Eqs.\,\ref{eq:p1d_polynomial},\ref{eq:p1derr_polynomial}) with the true values using a log-likelihood loss function $\mathcal{L}$ 
\begin{equation}
    \mathcal{L} = \sum_i^{N}\left[\sum_{k}^{N_k}\left(\frac{\log_{10} P_{\rm 1D}^{\prime\ \rm{pred}} -  \log_{10} P_{\rm 1D}^{\prime\ \rm{true}}}{\sigma_{\log_{10} P^\prime_{\rm 1D}}}\right)^2+2\log{\sigma_{\log_{10} P^{\prime}_{\rm 1D}}}\right] \cdot \frac{1}{N}\, ,
    \label{eq:loss_function}
\end{equation} where $N$ corresponds to the number of training samples considered in the loss and $N_k$ is the number of wave-number $k_{\parallel}$ bins. Evaluating the loss function in the \lc{$\log_{10}$} \poned space rather than the space of polynomial coefficients enables the neural network to learn the importance of each coefficient (e.g. the zero- and first- order polynomial degrees contribute more to the \poned than the fourth-order one) and weight the learning accordingly. In Appendix \ref{sec:pcaemu}, we also explore emulating the \poned in principal components space, but the polyfit emulator is more accurate and robust.

\section{Emulator characterisation}
\label{sec:characterisation}
In this section, we optimise the emulator's training strategy (\S\ref{sec:nn_characterisation}) and the training sample by studying whether the precision of the emulator increases after including optical-depth rescalings in the training sample (\S\ref{sec:rescalings}) and exploring different ways of doing data augmentation with the three-axes data (\S\ref{sec:trainingsample_characterisation}). \lc{To facilitate meaningful comparisons, we ensure that all the emulators examined in Figs.~\ref{fig:net_characterisation} and \ref{fig:sample_characterisation} share the same architecture, optimisation strategy, and initial random weights. This approach minimises the impact of randomness and provides a fair basis for evaluating their performance.}

\lc{In this paper, we will adopt the percent error as the primary metric for evaluating the performance of the emulator. The percent error is defined as }
\begin{equation}
    {\rm Percent\ error} = \left(\frac{P_{\rm 1D}^{\rm Pred}}{P_{\rm 1D}^{\rm True}} -1 \right) \cdot 100
    \label{eq:percent_error}
\end{equation}

\subsection{Neural network characterisation}
\label{sec:nn_characterisation}

\begin{figure}
\includegraphics[width= 0.48\textwidth]{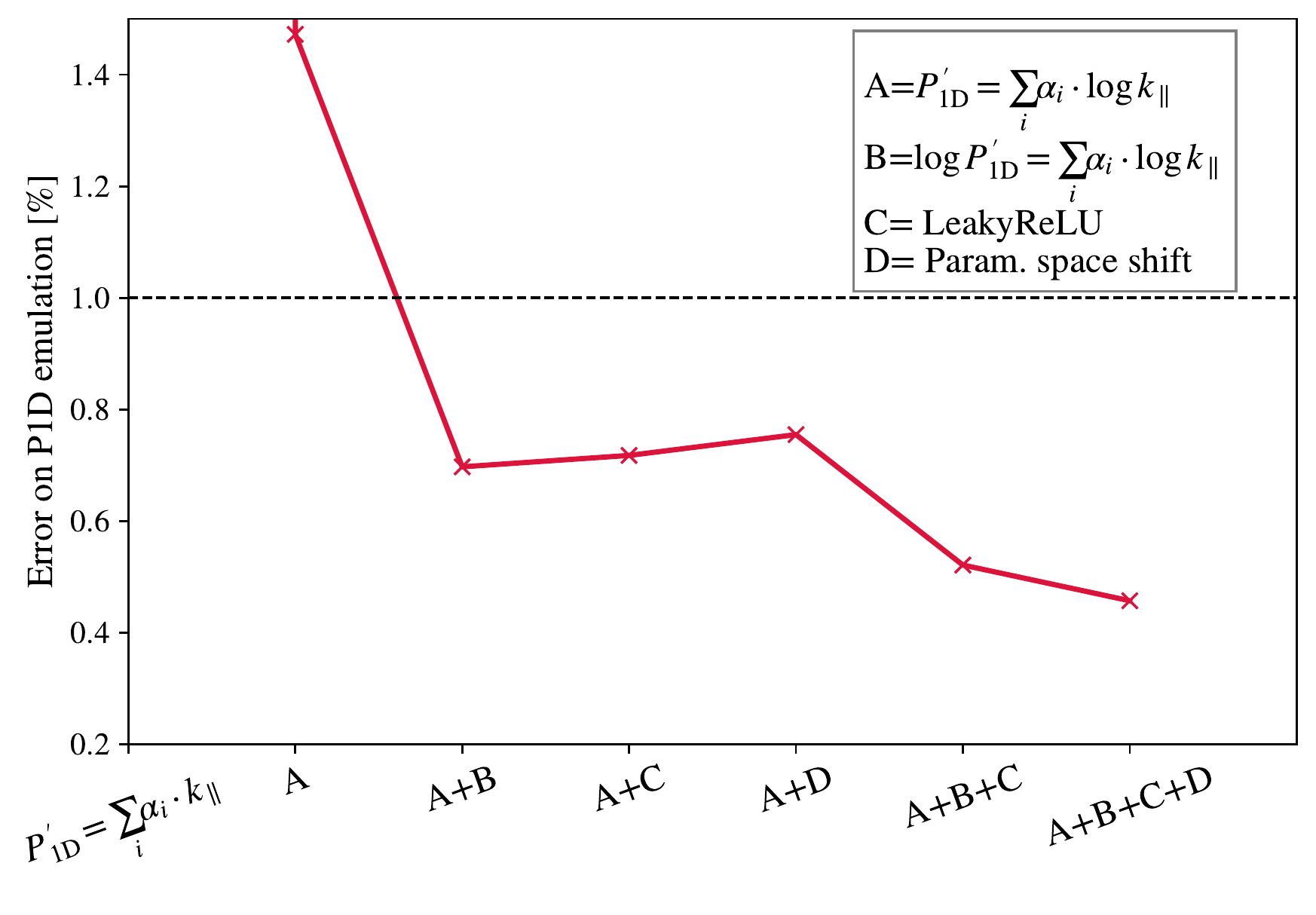}
\centering
\caption{Optimisation of the neural network emulator. The $x$-axis corresponds to different modifications implemented on the network's architecture, training data, and training procedure. The $y$-axis indicates the percent error in the \poned emulation for each of these modifications. \lc{The first point (not visible in the plot) corresponds to a 10\% measurement error.}}
\label{fig:net_characterisation}
\end{figure}

In order to optimise the performance of the emulator, we have tested several actions on the network architecture, the training procedure, the input data, and the target \poned. Figure \ref{fig:net_characterisation} shows the impact of some of these actions on the emulator's performance. In the $x$-axis, the plot indicates the action performed on the network, while the $y$-axis displays the percent error in the emulated \poned. Comparing the error after each modification, we can quantify the impact that each action has on the \poned. In all cases, the emulator is trained on the \lacehc simulation pairs (\S\ref{sec:methods_sims}) and tested on \simcentral. To reduce the sources of variability between runs, the neural network is initialised with the same set of initial weights in all runs, which are also randomly generated.

We first implemented one modification at a time to test their effect independently of the others, and then tested the impact of cumulative modifications. The starting point is a neural network fitting a polynomial in $k_{\parallel}$ and predicting \ponedprime (Eq.\ref{eq:p1d_scaling}). We first study the impact of fitting the \ponedprime in $\log_{10} k_\parallel$ space (A). This drastically reduces the error from 10\% to 1.4\%. \lc{Historically, the P1D has been described using powers of $k_{\parallel}$, as seen in e.g. Eq.~14 in \citet{p1d_PalanqueDelabrouille2013} and Eq.~4 in \citet{Karacayli2022}.} The second action (B), consists in fitting the $\log_{10}(\ponedprime)$ instead of \ponedprime directly, which clearly benefits the emulator.

Actions C and D modify the activation function and the input's normalisation, respectively. Initially, the input parameter space ($\theta$) was normalised with a min-max normalisation 
\begin{equation}
    \vec{\theta}^\prime = \frac{\vec{\theta}- \mathrm{min}(\vec{\theta})}{\mathrm{max}(\vec{\theta}) - \mathrm{min}(\vec{\theta})}\, ,
\end{equation}
which rescales the parameter space to the [0,1] range. Instead, we modify this scaling by also subtracting 0.5, in such a way that our input parameter space is the [-0.5,0.5] range. Since the ReLU activation function sets to zero all negative values, we include the Leaky-ReLU activation function \citep{LeakyReLU_Xu2015} to avoid vanishing gradients. In Fig.\ref{fig:net_characterisation}, we can see that the Leaky-ReLu alone does not have an impact on the performance (A+C), but it does when it is combined with the prediction in $\log_{10} k_\parallel$ (A+B+C). Similarly, the input parameter shift does (A+D) does not improve the emulator's performance but combined with the other changes there is a small improvement. 

Based on these results, the configuration of our fiducial emulator includes the fitting in $\log_{10} k_{\parallel}$, the fitting of the $\log_{10}$\,\poned, the Leaky-ReLU activation function and the parameter shift (A+B+C+D).

\subsection{Optical-depth rescalings}
\label{sec:rescalings}

\begin{figure}
\includegraphics[width= 0.48\textwidth]{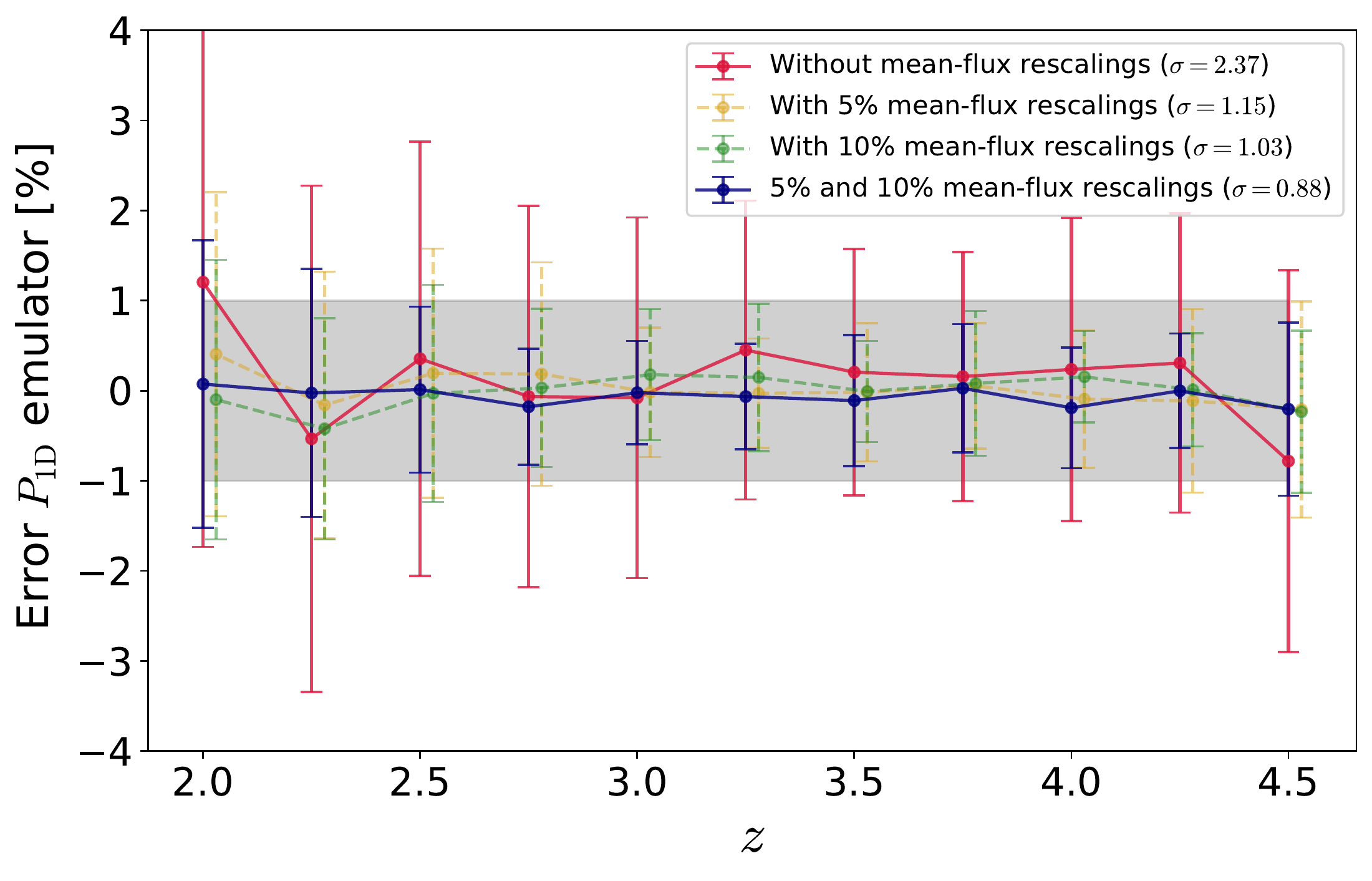}
\centering
\caption{Impact of the optical-depth rescalings on the \poned emulation. The plot shows the percent error in the \poned emulation for the 30 hypercube simulations pairs averaged over simulations and scales. The solid red and solid blue lines correspond to samples without and with all rescalings, respectively. In turn, the soft dashed lines indicate the samples including 5\% (yellow) and 10\% (green) optical-depth rescalings. The yellow and green lines are artificially shifted in the $x$ axis for a better visualisation.}
\label{fig:tau_rescalings}
\end{figure}

Generating hydrodynamical simulations and extracting the \poned is computationally expensive and time consuming. However, once the \poned measurement per snapshot has been measured, we can generate several realisations of such \poned rescaling the mean-transmitted flux fraction \citep{hydro_Lukic2015, hydro_Walther2021}. This is an effective way of increasing the training sample of the emulator. Our simulations include four mean-flux rescalings per snapshot, which augment the training sample from 330 to 1650 training points. Such rescalings correspond to $\pm$(5 and 10)\% in optical depth, which does not propagate into $\pm$(5 and 10)\% mean flux. This is a novelty of the \texttt{LaCE-NN} emulator that can benefit from the training sample augmentation with optical-depth rescalings.

Figure \ref{fig:tau_rescalings} compares the percent error in the \poned predictions with emulators trained on samples without optical-depth rescalings (red), with 5\% rescalings (blue), 10\% (green), and combining both 5\% and 10\% optical-depth rescalings (yellow). For this plot, we have run four independent leave-one-out tests (see \S\ref{sec:L1O}), on per training set, obtaining the error in the \poned emulation of each simulation and snapshot for each of the training samples. The error is then averaged over simulations and scales to have an error estimate per redshift. The emulator is always tested on non-scaled \poned.

There is a significant improvement in the performance of the emulator when rescalings are incorporated into the training sample. For instance, the inclusion of 5\% optical-depth rescalings resulted in a reduction of the percentage error in the emulated \poned by a factor of two. The emulator benefited more from the inclusion of 10\%  optical-depth rescalings as compared to the 5\% ones (green vs yellow). This may be due to the fact that the 10\% rescalings increase the parameter space to a greater extent than the 5\% rescalings, leading to the new parameter space encompassing the 5\% mean-flux rescaling points as well. 

Nevertheless, the optical-depth rescalings included in this post-processings mostly populate the parameter space and barely affect the limits of its convex hull. Future emulators aiming to predict universes with different IGM history will also require temperature and mean-flux rescalings increasing the covered parameter space.

\subsection{Training sample characterisation}
\label{sec:trainingsample_characterisation}

\begin{figure}
\includegraphics[width= 0.5\textwidth]{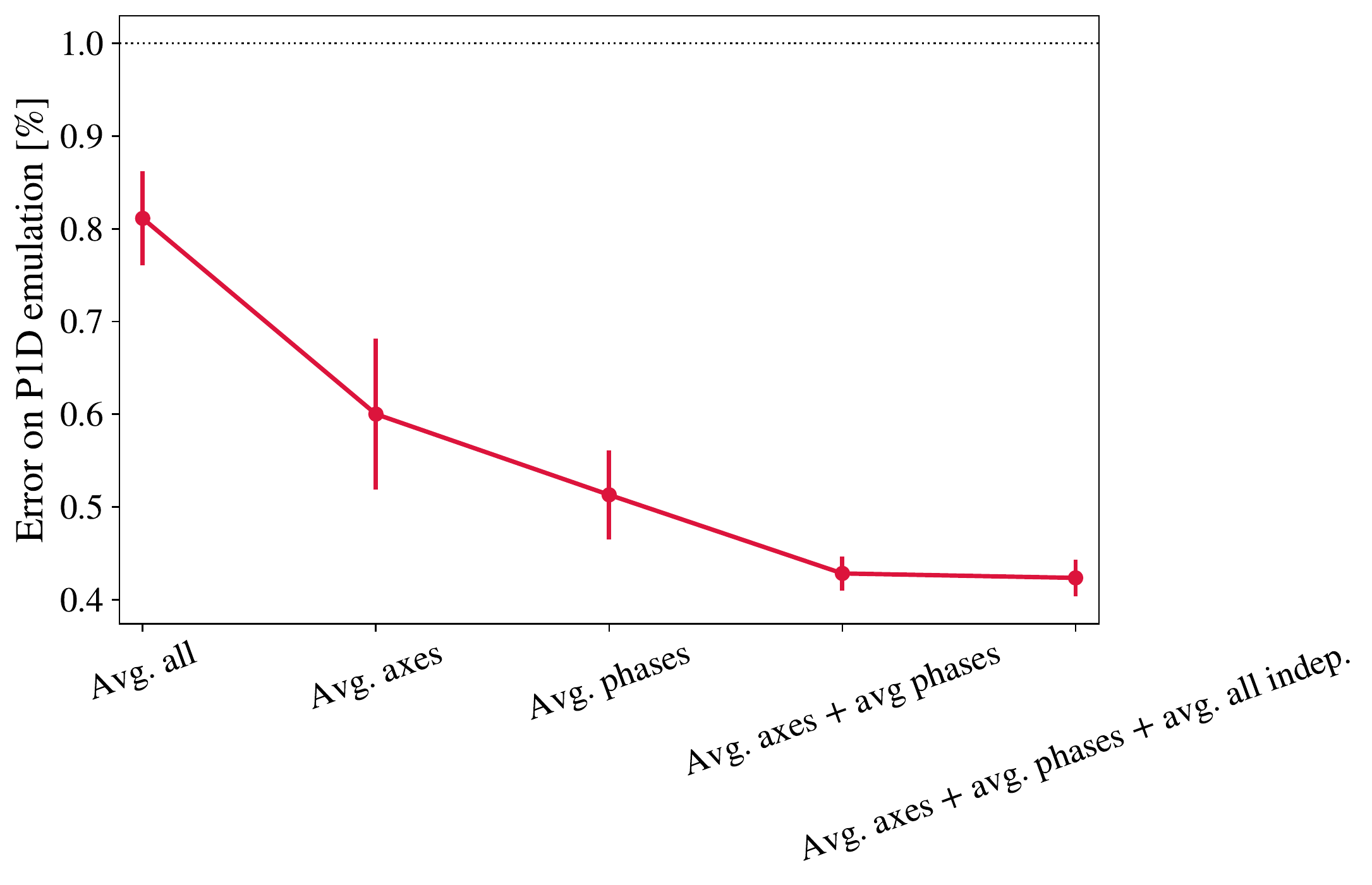}
\centering
\caption{Characterising the training sample of the emulator. The figure shows the percent error obtained in the \poned emulation for different training sample constructed from several possible combinations of \poned measurements over the three axes and two paired phases. The horizontal lines correspond to  the 1\% error requirement (dotted-black line) and the percent error of the one-axis emulator (blue lines). \lc{The error bars are derived from the dispersion of the percent error obtained from repeating the measurement ten times. The different training samples have been tested on the \simcentral simulation.}}
\label{fig:sample_characterisation}
\end{figure}

With the new post-processing of the hydrodynamical simulations, we obtain one \poned measurement along each of the three axes for every simulation box. Given that our simulations are fixed-and-paired (\S\ref{sec:methods_sims}), we have a total of six \poned measurements per simulation snapshot. These measurements can be averaged to obtain a single, less noisy \poned measurement.

However, in this section, we explore data augmentation techniques by considering different combinations of these six \poned measurements. While the most common approach involves training the emulator with the average of the six measurements, we also investigate the possibility of utilising each measurement independently and combining only \poned along axes or phases.

Figure \ref{fig:sample_characterisation} explores the emulator's performance by considering different combinations of the six \poned measurements. The plot depicts the percent error in \poned emulation for various training sets. These errors were estimated using all \lacehc simulations and are compared to the most accurate \poned measurement available, which results from averaging all axes and phases.

The red line represents the percent error in \poned emulation for different training samples. The first red point, labeled "Avg. all," corresponds to a training sample that averages the \poned measurements from all axes and phases, resulting in a total of 1650 training points (30 simulations $\times$ 11 snapshots $\times$ 5 optical-depth rescalings). 

The next red points are ``Avg. axes'' and ``Avg. phases'', and reduce the percent error compared to the previous result. In the first case, ``Avg. axes'', we have averaged the \poned from over the three axes, but not over phases. Therefore, this corresponds to two \poned measurements per snapshot. Additionally, we also include the \poned averaged along axes and phases, which is the most accurate measurement we have. Therefore, this corresponds to 4950 training points (30 simulations $\times$ 11 snapshots $\times$ 5 optical-depth rescalings $\times$ (2+1) \poned). In the second case, ``Avg. phases'', we average over phases, but not over axes. For each snapshot, we have three \poned measurements, and as before, we additionally include the average over axes and phases. Therefore, the training sample contains 6600 training points (30 simulations $\times$ 11 snapshots $\times$ 5 optical-depth rescalings $\times$ (3+1) \poned). 

The following red point, ``Avg. axes + Avg. phases'' combines the two previous ones, which makes a total of 9900 training points (30 simulations $\times$ 11 snapshots $\times$ 5 optical-depth rescalings $\times$ (3+2+1) \poned) and clearly reduces the percent error compared to using only one of the data sets. Finally, the last point, ``Avg. axes + Avg. phases + indep'' adds up the average over axes, the average over phases and the six \poned measurements, which correspond to 19800 training points (30 simulations $\times$ 11 snapshots $\times$ 5 optical-depth rescalings $\times$ (3+2+1+6) \poned). In this case, the emulator does not seem to benefit from the additional training points. 

Based on these results, the fiducial training sample for our emulator corresponds to ``Avg. axes + Avg. phases'' and contains 9900 training points.


\section{Testing emulation strategy}
\label{sec:results}

In this section, we evaluate the performance of the proposed emulation approach for the surrogate models discussed in the preceding sections. We specifically focus on assessing the precision of the emulator when applied to \lacehc simulations (\S\ref{sec:L1O}) and redshifts (\S\ref{sec:other_redshift}) that were not part of the training sample. In \S\ref{sec:other_comparison}, we compare the \texttt{LaCE-GP} and the \texttt{LaCE-NN} emulators. Additionally, we investigate the effect that cosmic variance has on the emulator (\S\ref{sec:other_cvariance}) and the emulator's performance on cosmologies (\S\ref{sec:other_cosmo}) and IGM thermal histories (\S\ref{sec:other_astro}) that were not included in the training set.


\subsection{Leave-one-out tests}
\label{sec:L1O}

\begin{figure}
\includegraphics[width= 0.48\textwidth]{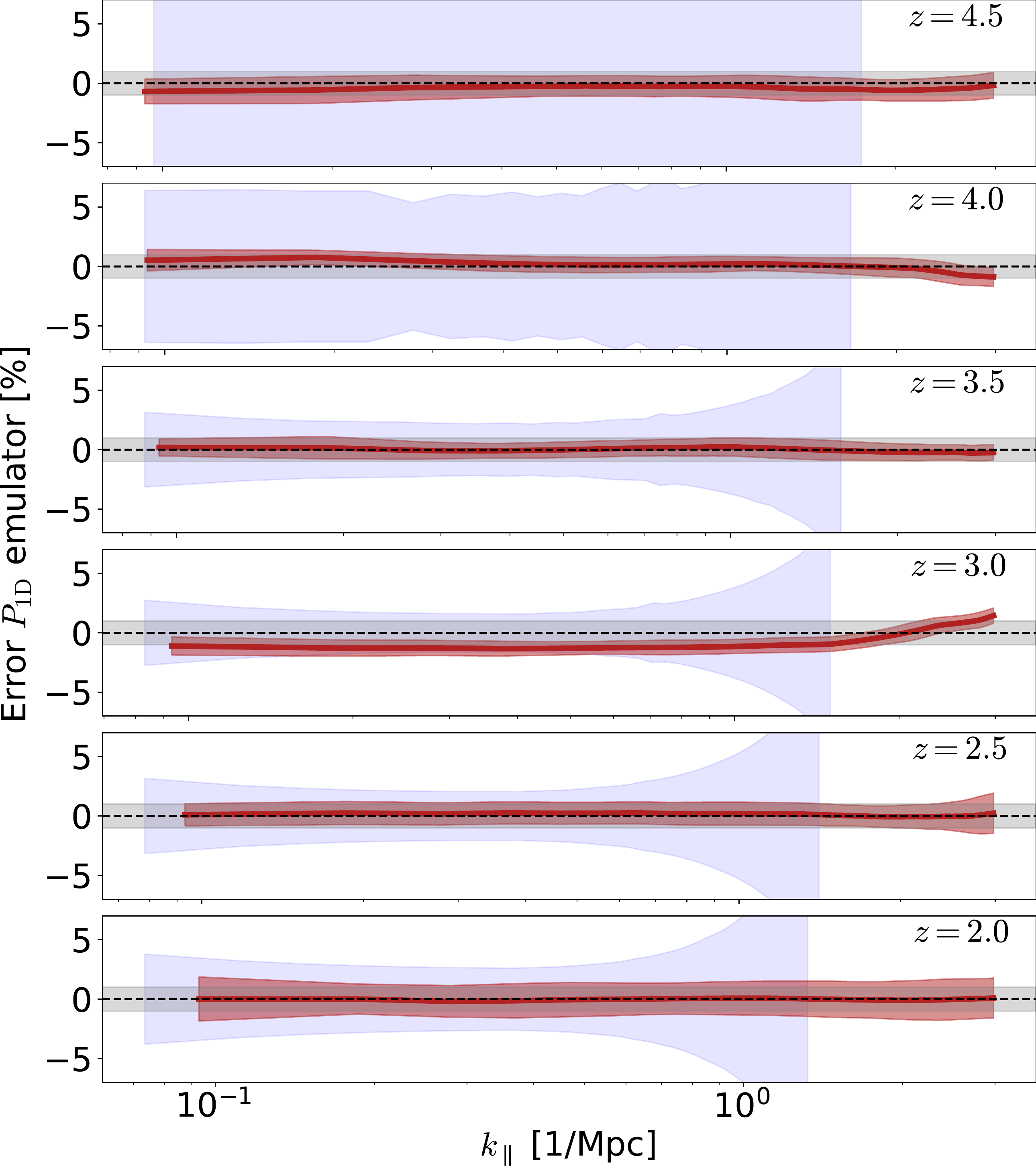}
\centering
\caption{Leave-one-out test. Percent error in the emulator \poned at different redshifts. The shaded grey area indicates the 1\% error requirement for the emulator. The red line show the mean percent error and the red shaded region corresponds to the standard deviation of the percent error. The blue shaded region represents the observational uncertainties from the latest analysis of BOSS/eBOSS data. }
\label{fig:leave_one_out}
\end{figure}

The leave-one-out test is a commonly used validation test to evaluate the accuracy of an emulator. The emulator is trained using the set of training simulations, holding one out as a validation set, which is then evaluated. The leave-one-out test is repeated for each simulation in the training set, determining how well the emulator is able to generalise to unseen simulations and the overall accuracy it can achieve.

Figure \ref{fig:leave_one_out} presents the leave-one-out test results for \lacehc simulations. To generate the plot, we have optimised 30 independent emulators, each one training on 29 \lacehc simulation pairs and evaluated on the left-out simulation. 

To produce the Fig.~\ref{fig:leave_one_out} we group all snapshots with the same redshift and take the mean and the standard deviation of the percent error in the emulation across simulations.  The red line indicates the mean percent error and the red shaded region corresponds to the standard deviation of the percent error. The shaded grey area in the figure indicates the 1\% error requirement for the emulator. For most cases, the emulator reaches the $<1\%$ error requirement at all redshifts, although some measurements at $z=2$ are slightly over 1\% error. This is not unexpected since $z=2$ lies in the extreme of the \lyaf detection and therefore the emulator is more likely to degrade. Finally, the blue-shaded region represents the observational uncertainties from the latest analysis of BOSS/eBOSS data \citep{p1d_Chabanier2019}.


\subsection{Redshifts outside the training set}
\label{sec:other_redshift}

\begin{figure}
\includegraphics[width= 0.48\textwidth]{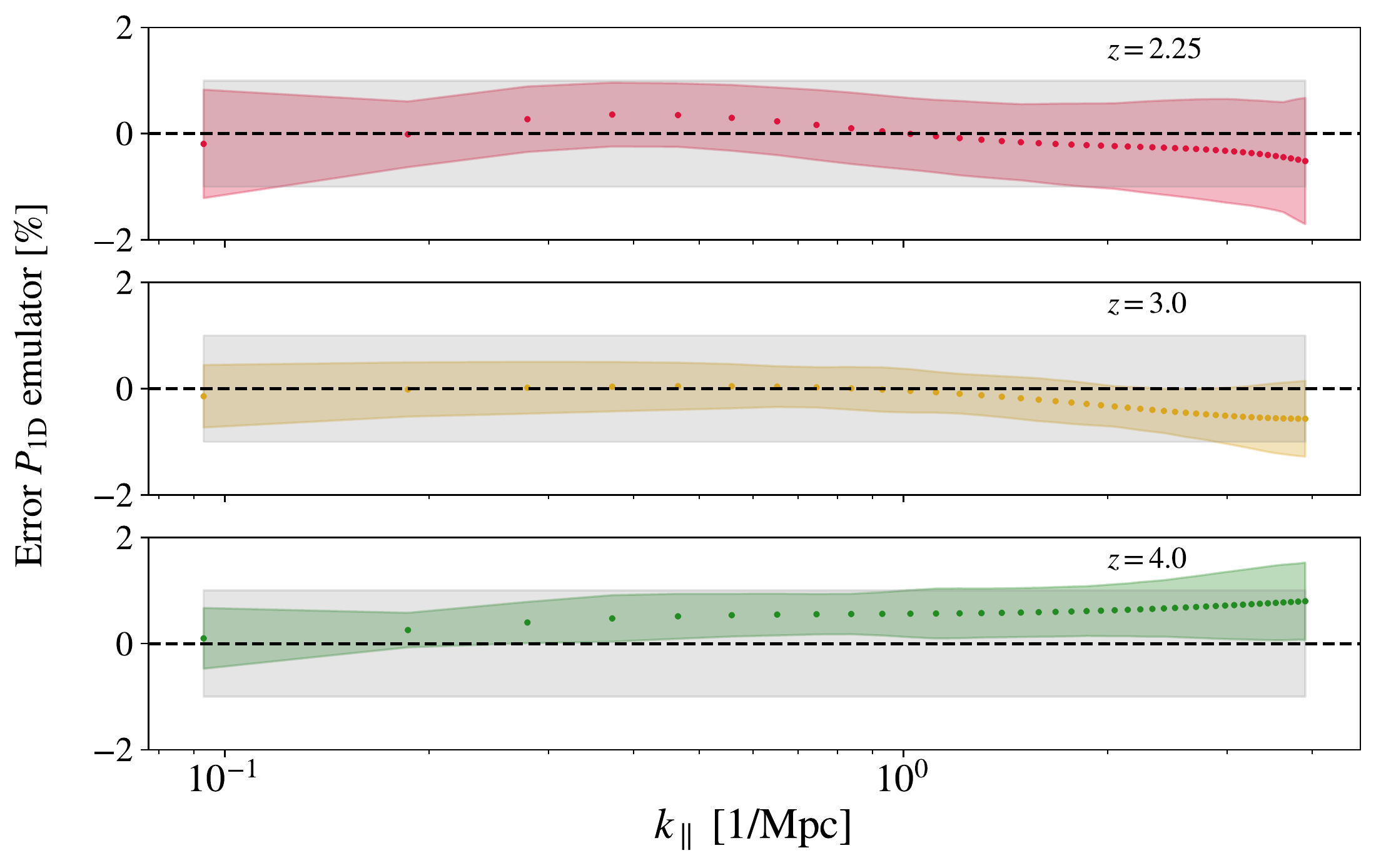}
\centering
\caption{Emulator's performance at arbitrary redshifts. The plot shows the percent error in the \poned emulation at redshifts $z=2.25$, $z=3$, $z=4$, when all snapshots at these redshifts are dropped from the training sample. The dotted line indicates the mean percent error, while the shaded-coloured area corresponds to the standard deviation across simulations. The shaded-gray area indicates the 1\% error requirement. We specifically choose to test the emulator at $z=2.25$ instead of $z=2$ due to the latter being on the boundary of the convex hull. Consequently, if we were to exclude $z=2$ from the training sample, it would no longer be within the training space.}
\label{fig:other_redshift}
\end{figure} 

\lc{The \texttt{LaCE} emulators do not consider redshift as an input parameter, allowing it to make predictions on \poned values for redshifts that were not part of the training sample. By constructing the emulator using training points from all redshifts and assuming that the \poned can be adequately described by the six parameters defined in the emulation strategy (\S\ref{sec:methods_params}), the need to explicitly include redshift information in the emulator is eliminated. This flexibility enables the analysis of data with any desired redshift binning.}

\lc{To test if the assumption holds and the \poned is fully described by the training parameters without explicitly including redshift information,} we have evaluated the emulator's performance at a given redshift $z=z_0$, when all snapshots at that redshift are removed from the training sample. In the top, middle, and bottom panels of Fig. \ref{fig:other_redshift}, we show the percent error in the emulated \poned at $z=2.25$, $z=3$, and $z=4$, respectively, when the corresponding testing redshift is removed from the training sample. The shaded grey area indicates the 1\% error requirement, which is achieved in all cases. This proves that the emulator can make \poned predictions at arbitrary redshifts. We specifically choose to test the emulator at $z=2.25$ instead of $z=2$ due to the latter being on the boundary of the convex hull. Consequently, if we were to exclude $z=2$ from the training sample, it would no longer be within the training space. Therefore, to ensure a comprehensive evaluation, we opt to test the emulator at $z=2.25$ instead.


\subsection{Emulator comparison}
\label{sec:other_comparison}

The success of this initial version demonstrated that we could accurately emulate the \poned to sub-percent errors from the amplitude and slope of the linear matter power spectrum. However, this first emulator version was only tested on simulations with the same IGM history as the training set. 

\texttt{LaCE-NN} incorporates several modifications compared to \texttt{LaCE-GP}. The most notable difference is the replacement of Gaussian processes with neural networks. Additionally, \texttt{LaCE-NN} incorporates a new post-processing step that measures the \poned along all three simulation axes (\S\ref{sec:methods_post}) using 768 skewers. This increases the resolution of the measured \poned and provides three \poned measurements per snapshot. Furthermore,  since our simulations are fixed-paired simulations, we have six \poned measurements per snapshot. The new post-processing step also includes applying five optical-depth rescalings to each snapshot (\S\ref{sec:rescalings}), resulting in a total of 9900 \poned measurements. This represents a substantial improvement over the \texttt{LaCE-GP} data processing, which yielded 330 \poned measurements, and already produces datasets that are difficult to handle with GPs.

\subsection{Cosmic variance}
\label{sec:other_cvariance}

Even though the \lacehc simulations use ``fixed-and-paired'' conditions, their volume is very limited and thus cosmic variance should affect emulator predictions significantly. To estimate the impact of cosmic variance on the results, we compare the performance of the emulator for the \simcentral and \simseed simulation pairs (see \S\ref{sec:methods_sims}). These pairs present cosmological parameters at the centre of the \lacehc parameter space; therefore, the precision of the emulator should be close to optimal for them. On the other hand, the \simcentral pair use the same set of Fourier phases as the \lacehc simulations, while the \simseed simulations use different initial conditions. As a result, any difference in the performance of the emulator for these simulations isolates the impact of cosmic variance.

In the first two panels of Fig.~\ref{fig:other_cosmo}, we display the precision of the \texttt{LaCE-GP} and \texttt{LaCE-NN} emulators for these two simulation pairs. As we can see, the two emulators present better than 1\% performance for both cases, confirming that the impact of cosmic variance on emulator predictions is negligible. This is the consequence of using a polynomial fit to smooth \poned before training and testing the emulator (see Appendix~\ref{sec:poly_order}), which is justified by the smoothness of this observable and greatly reduces the impact of cosmic variance on large scales. As shown in \citet{pedersen2023CompressingCosmologicalInformation}, the impact of cosmic variance is much larger if no smoothing is applied (see also Appendix~\ref{sec:pcaemu} for another smoothing strategy).


\subsection{Cosmologies and $\Lambda$CDM extensions not considered in the training set}
\label{sec:other_cosmo}

\begin{figure}
\includegraphics[width= 0.48\textwidth]{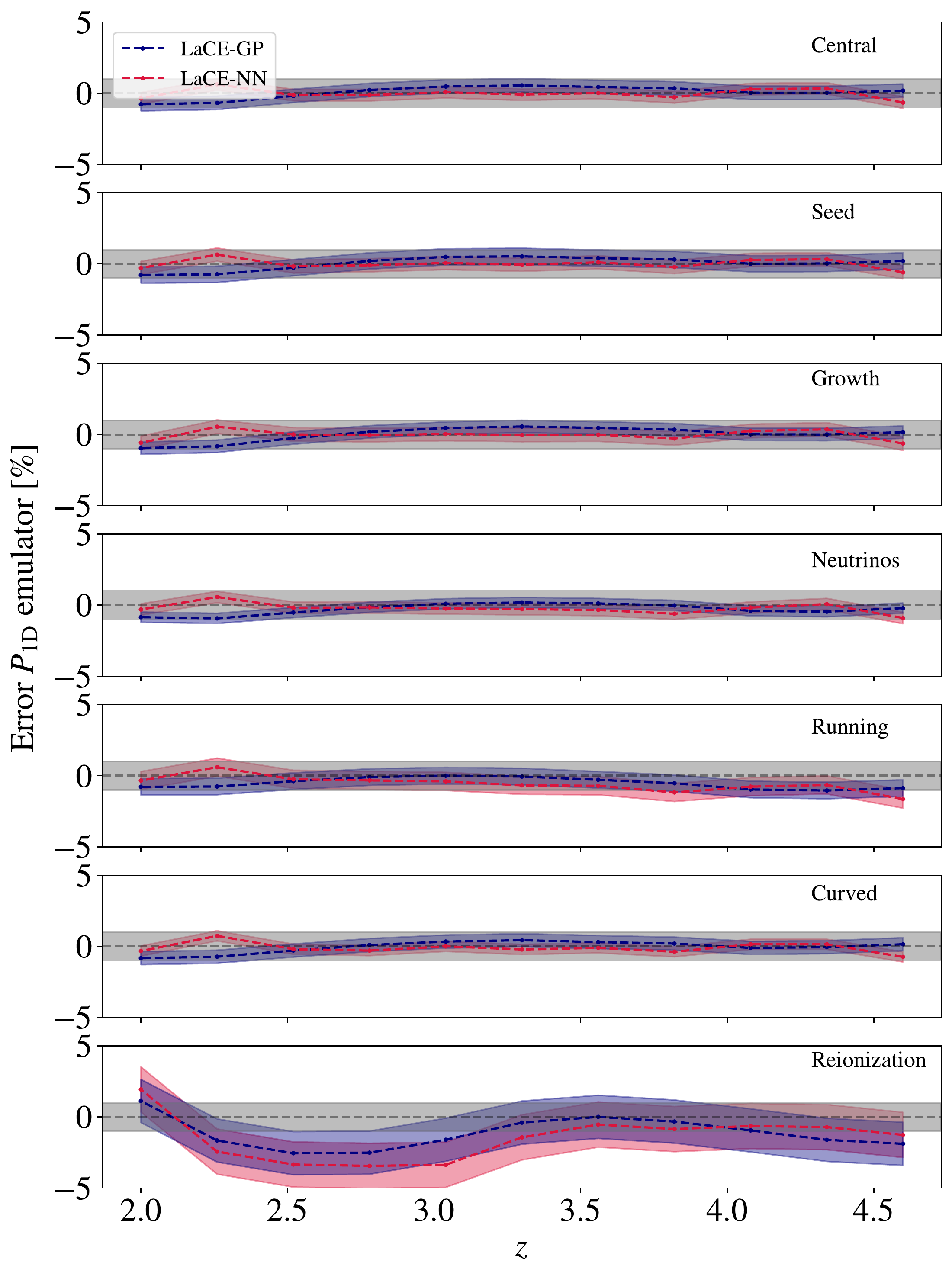}
\centering
\caption{Emulator performance on simulations with different cosmologies as the one sampled by the \lacehc simulations. The $\nu$-sim (top panel) contains massive neutrinos, while the $h$-sim (bottom panel) modifies the growth rates with respect to the training sample. The dotted line corresponds to the mean percent error in the \poned prediction, while the shaded area indicates the standard deviation across $k_{\parallel}$.}
\label{fig:other_cosmo}
\end{figure}

The \lacehc simulations adopt a standard $\Lambda$CDM parameterisation and explore different values of the amplitude and slope of the primordial power spectrum while considering the same expansion and growth histories (see \S\ref{sec:methods_sims}). Given that \poned is sensitive to the velocity field, it is important to check the precision of the emulator for other growth rates. To do so, we analyse the \simh simulation pair (see \S\ref{sec:methods_sims}), which uses a different value of $h$ relative to the \lacehc simulations. In the third panel of Fig.~\ref{fig:other_cosmo}, we show that our emulators present sub-percent precision for this pair, confirming that our emulation strategy enables producing precise predictions for cosmologies outside the training set \citep{Pedersen2021, pedersen2023CompressingCosmologicalInformation}.

We now proceed to study if the emulation strategy also works for three $\Lambda$CDM extensions: massive neutrinos, running of the spectral index, and curvature. We test the previous scenarios using the \simnu, \simns, and \simcurved simulation pairs (see \S\ref{sec:methods_sims}), which present extreme values for the previous ingredients already ruled out by observations \citep{planckcollaboration2020Planck2018Resultsa}: $\sum m_{\nu}=0.3$ eV neutrino mass, $\mathrm{d}\,n_s/\mathrm{d}\log k=0.015$, and $\Omega_k=0.03$, respectively. Remarkably, our emulators also present sub-percent precision for these simulations, as shown in the fourth, fifth, and sixth panels of Fig.~\ref{fig:other_cosmo}. Consequently, we can use this emulation strategy to set accurate constraints on $\Lambda$CDM extensions.


\subsection{Thermal histories not included in the training set}
\label{sec:other_astro}

\begin{figure}
\includegraphics[width=\columnwidth]{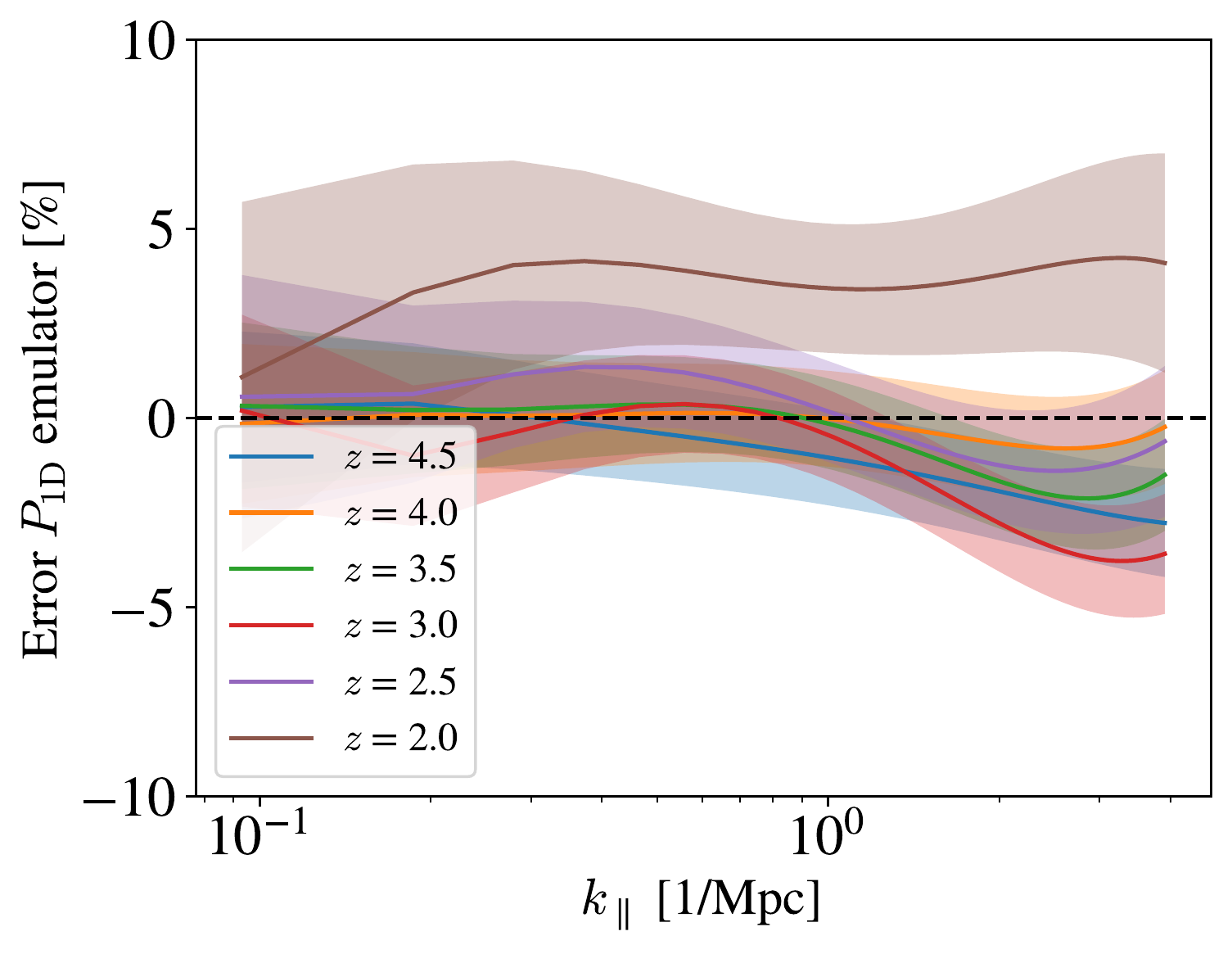}
\centering
\caption{Percent error in the \simigm emulation for snapshots at different redshift. The solid lines indicate the mean percent error while the shaded areas correspond to the emulator's predicted uncertainty. The prediction at $z=2$ shows worse performance than the rest of the snapshots, which is potentially because at such redshift, \mflux and $T_0$ are outside the convex hull of the training data.}
\label{fig:P18k4}
\end{figure}

\begin{figure}
\includegraphics[width=\columnwidth]{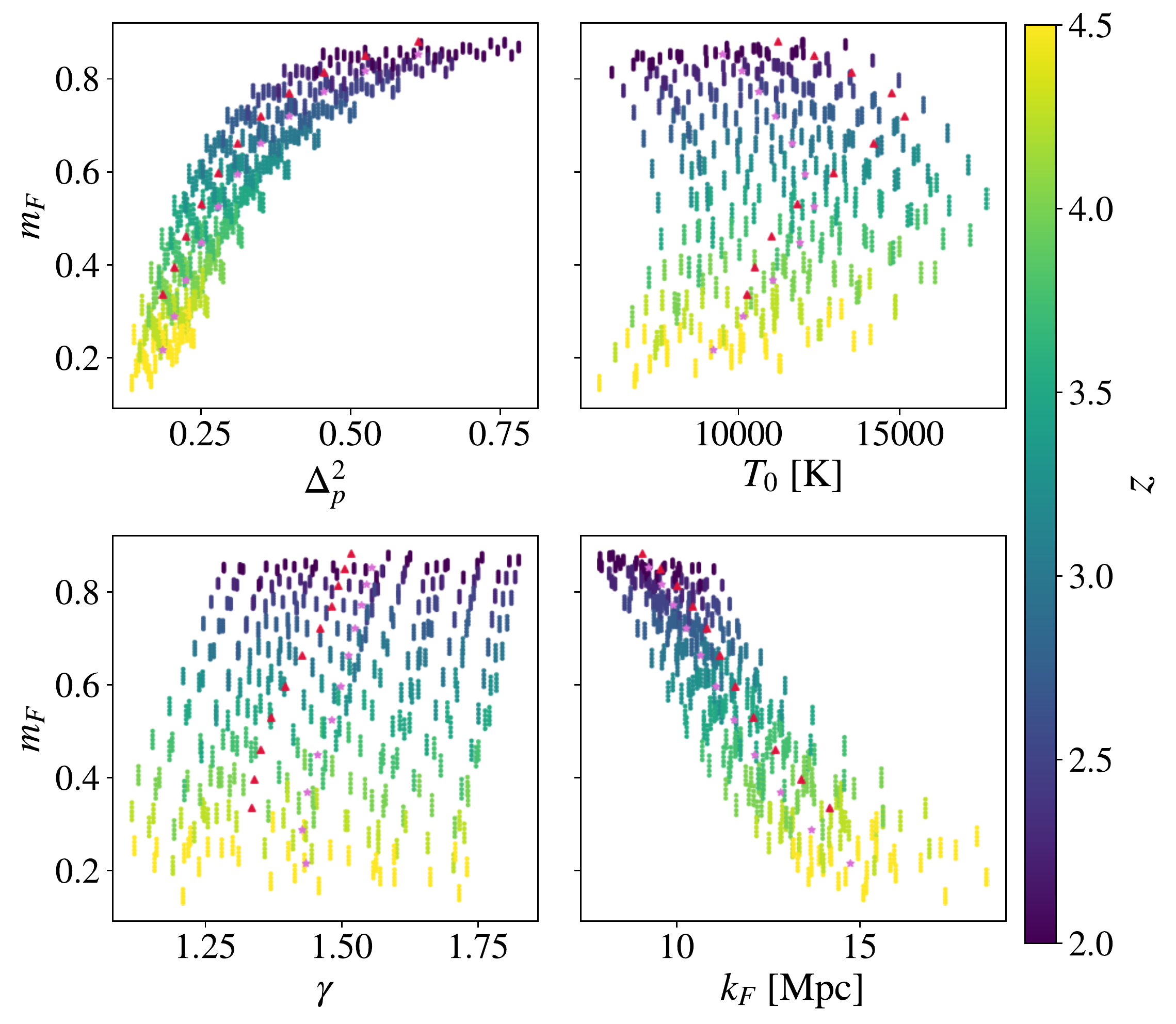}
\centering
\caption{Distribution of \lacehc, \simcentral, and \simigm simulations in the space of emulation parameters. Small dots indicate the results for \lacehc simulations with optical depth rescalings, while pink stars and red triangles show the results for \simcentral and \simigm, respectively. Small dots are coloured by the redshift of the simulation snapshot for visual purposes, as the redshift information is not considered by the emulator. Even though the \simigm and \lacehc simulations present different astrophysical implementations, we can readily see that the first lies within the range of the parameter space covered by the latter.}
\label{fig:dist_params}
\end{figure}

In \S\ref{sec:methods_sims}, we introduced the \simigm pair, which implements a different \ion{He}{ii} reionisation history than the \lacehc simulations, which result in a very different thermal history (see Fig.~\ref{fig:T0_vs_z}). In the last panel of Fig.~\ref{fig:other_cosmo}, we show the performance of the \texttt{LaCE-GP} and the \texttt{LaCE-NN} emulators for this pair. As we can see, the precision of both emulators is similar and on average better than 1.5\%, letting us conclude that the emulation strategy works even for \ion{He}{ii} reionisation histories not considered in the training set. This is the consequence of emulating in the $\sigma_T$-$\gamma$ space, which captures the state of the IGM at a particular cosmic time, instead of as a function of parameters encoding information about the entire ionisation or thermal history. 

We find that the overall precision of the emulator for this pair is on average four times worse than for the \simcentral pair, especially for low redshift. To further understand this decrease in performance, in Fig.~\ref{fig:P18k4} we display the results of the \texttt{LaCE-NN} emulator for the \simigm pair without collapsing the wavelength information. Lines indicate the results for different redshifts, while shaded areas denote the level of uncertainty predicted by the emulator. As we can see, the emulator produces biased results for $z=2$. To trace the origin of this issue, in Fig.~\ref{fig:dist_params} we display how the \lacehc, \simcentral, and \simigm simulations sample the emulator parameter space. Although the \simcentral and \simigm simulations present the same cosmology, we can readily see that the relation between \mflux and $T_0$ is very different for these. Furthermore, we find that the value of the \simigm parameters for $z=2$ lies outside of the parameter space covered by the \lacehc simulations, which explains the decrease in accuracy.

An important observation is that the emulator assigns a higher uncertainty to the \poned measurement at $z=2$. This increased uncertainty arises from the awareness that the parameters associated with $z=2$ lie outside the boundaries of the convex hull. By acknowledging this discrepancy, the emulator correctly accounts for the greater uncertainty in predicting \poned values at this particular redshift. 

Interestingly, the overall performance of the emulator for the \simigm pair is worse than for cosmologies and $\Lambda$CDM extensions not included in the training set. This result emphasises the significance of exploring various IGM models instead of only concentrating on densely sampling the cosmological part of the emulator parameter space. While some of this exploration can be carried out in a post-processing stage via mean flux and temperature rescalings \citep{hydro_Lukic2015}, the gas pressure smoothing scale cannot be rescaled so it is necessary to run simulations with different reionisation histories.


\section{Extended emulator}
\label{sec:extended_emulator}
\begin{figure*}
\centering
\includegraphics[width= 0.45\textwidth]{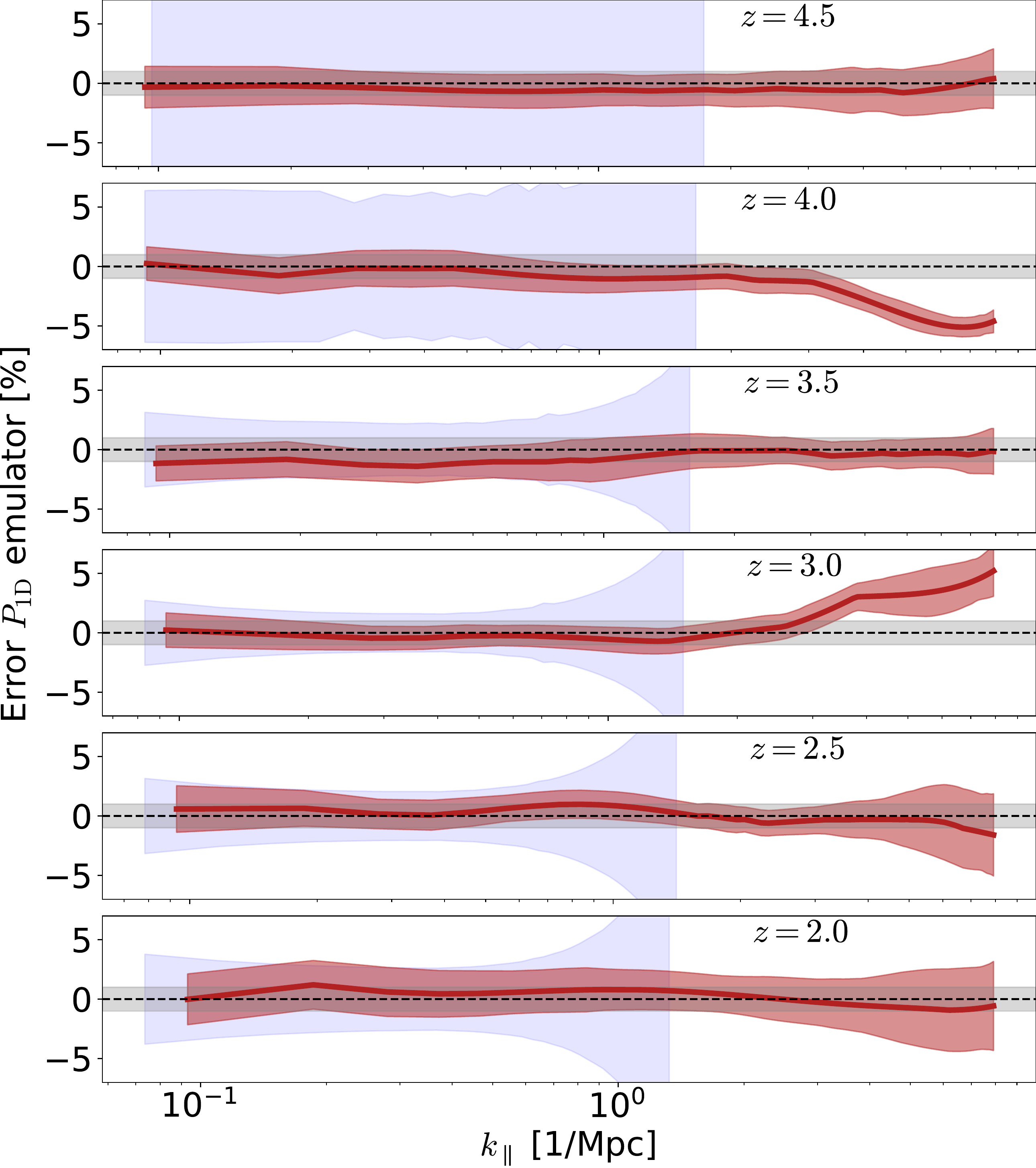}
\includegraphics[width= 0.45\textwidth]{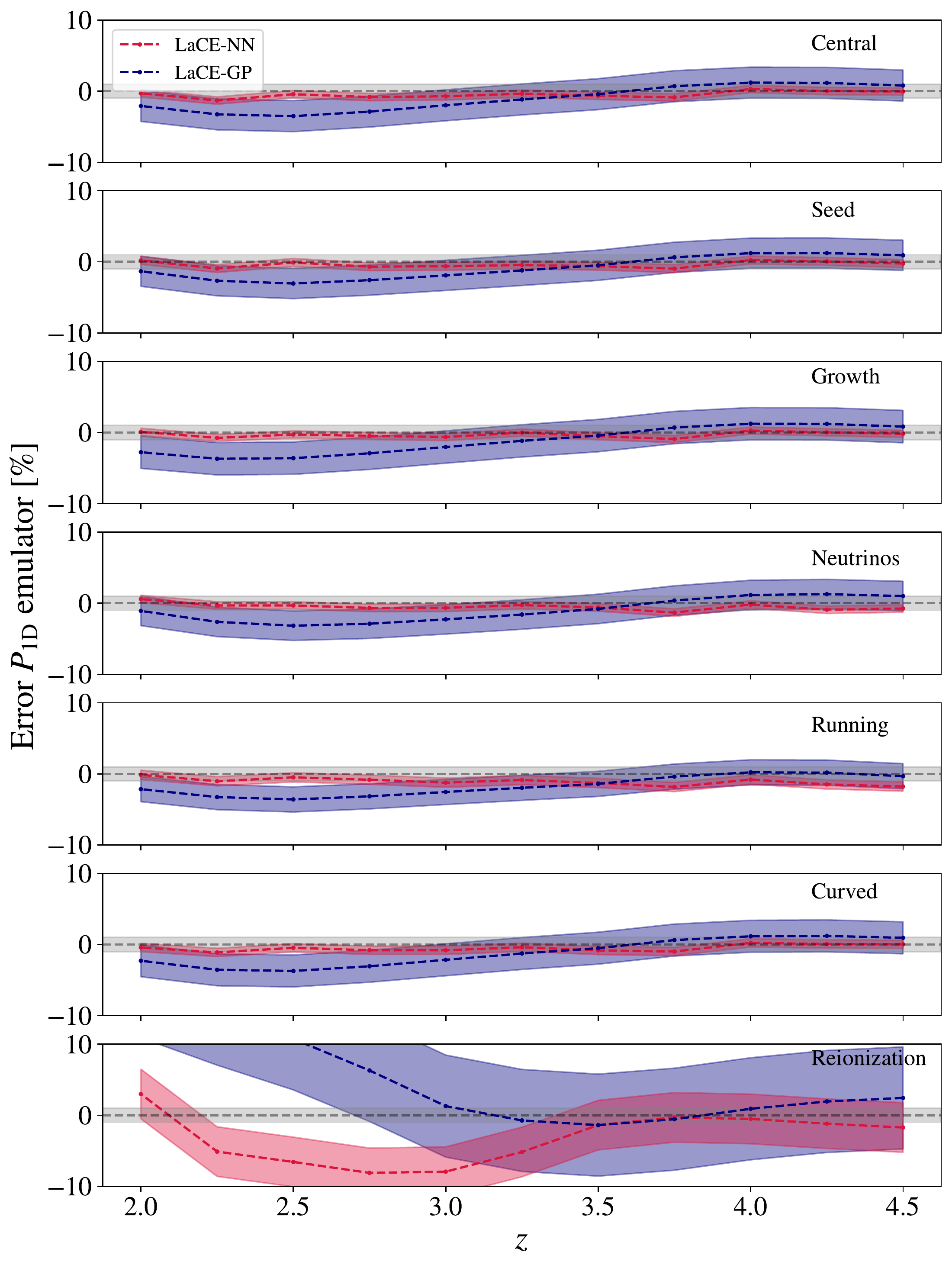}
\caption{ Performance of an extended version of the emulator to $k_{\parallel
}=8\ {\rm Mpc}^{-1}$ for \texttt{LaCE-NN} and \texttt{LaCE-GP}. The \emph{left} panel presents the leave-one-out test for the extended emulator. In the \emph{right}, the emulator is trained using the \lacehc simulations and tested on the seven test simulation pairs presented in \S\ref{sec:methods_sims}}
\label{fig:extended_emulator}
\end{figure*}

\begin{figure}
\includegraphics[width=\columnwidth]{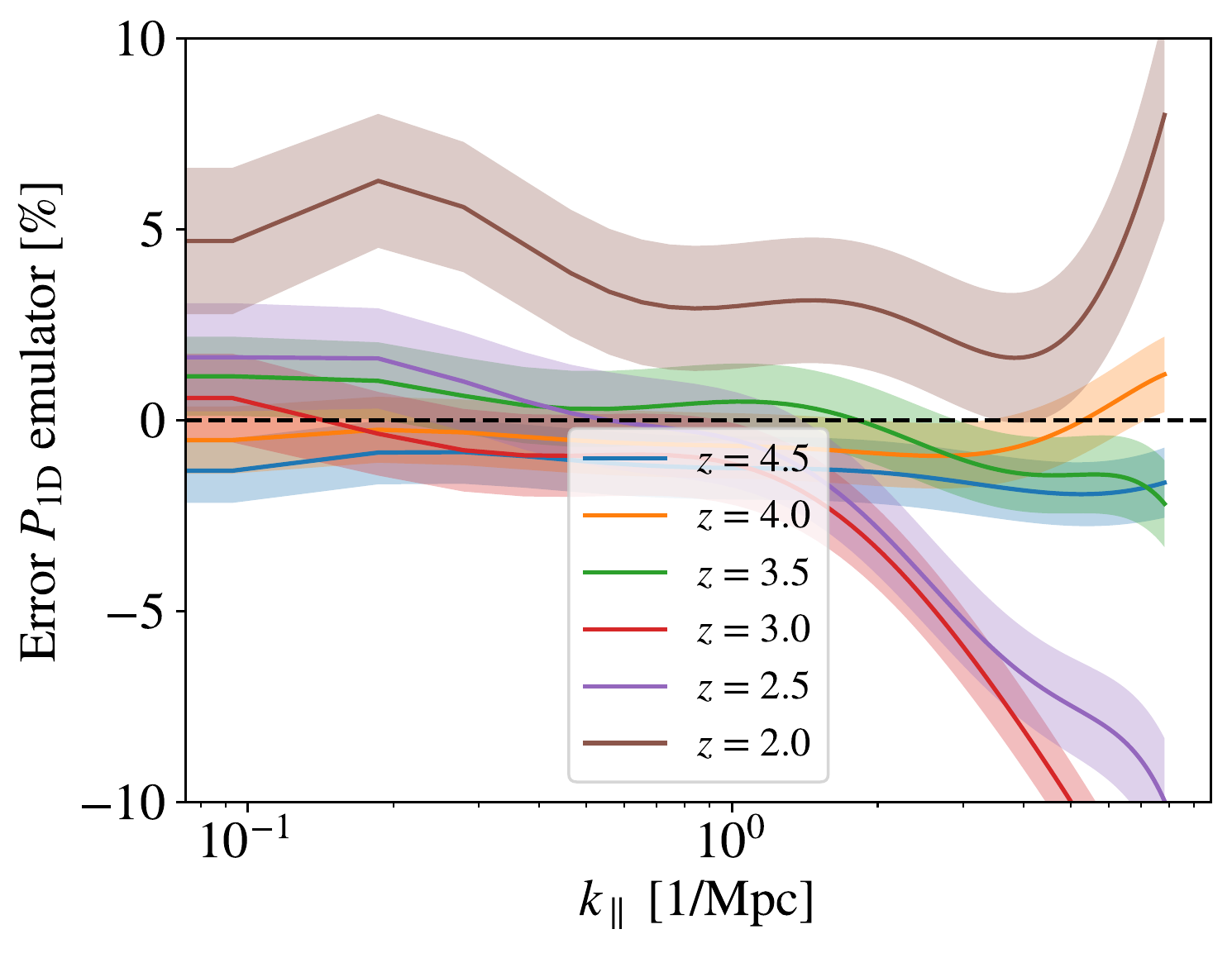}
\centering
\caption{Percent error in the \simigm emulation for snapshots at different redshifts. The solid lines indicate the mean percent error while the shaded areas correspond to the emulator's predicted uncertainty. This plot is analogous to Fig.~\ref{fig:P18k4} but for the extended emulator.}
\label{fig:P18k8}
\end{figure}

The \texttt{LaCE-GP} emulator has been developed and tested on scales ranging from $k_{\parallel
} \in (0,3] \ {\rm Mpc}^{-1}$. However, in order to cover the entire range of scales that DESI is designed to observe \citep{Ravoux2023,karaçayli2023} the \texttt{LaCE-NN} emulator has been developed to extend this range to $k_{\parallel} \in (0,4] \ {\rm Mpc}^{-1}$.  In this section, we present the results of an extended version of the \texttt{LaCE-NN} emulator, which predicts the \poned on scales up to $k_{\parallel}  = 8 \ {\rm Mpc}^{-1}$. It is worth noting that while DESI cannot observe scales beyond $k_{\parallel
} = 4 \ {\rm Mpc}^{-1}$, this extension has been motivated by the \poned measurements obtained from high-resolution quasars in \citet{Karacayli2022}.

Figure \ref{fig:extended_emulator} shows the performance of the extended emulator in the test simulations (left panel) and the \lacehc simulations (right panel). The figure includes the \texttt{LaCE-GP} emulator's performance for comparison purposes. In both cases, we employ the same emulator as used for $k_{\parallel
}<4\ {\rm Mpc}^{-1}$ without any adjustments to the hyperparameters. However, we modify the order of the polynomial used to fit the \poned, which is now set to $n=7$. Further details can be found in Appendix \ref{sec:poly_order}.

The \texttt{LaCE-NN} emulator outperforms the \texttt{LaCE-GP} performance in the extended case. Without any tuning of the emulator, \texttt{LaCE-NN} reaches a 1-2\% error in all test simulations but the \simigm one, which is the most challenging.  On the \lacehc simulations (right panel), \texttt{LaCE-NN} emulates the \poned of the \lacehc simulations with a mean percent error of only $\sim 2\%$. This is lower at $k_{\parallel
}<4\ {\rm Mpc}^{-1}$ and higher for smaller scales.  However, note that the error requirement for very small scales is not as stringent as $1\%$ \citep{Karacayli2022}. We present our initial attempt to create an extended emulator, acknowledging that there is room for improvement in future iterations. For instance, we aim to address the degradation observed at larger scales, and we propose potential enhancements such as down-weighting the contribution of smaller scales.

The extended emulator's performance exhibits a notable decline in the case of the \simigm simulation (Fig.~\ref{fig:P18k8}). Specifically, we observe a significant degradation in performance compared to the fiducial emulator. The errors primarily stem from inaccuracies at high $k_{\parallel}$ values, but we also observe a deterioration in predictions across all scales, particularly at $2<z<3$. At the redshifts in question, the parameter values for \mflux and $T_0$ lie at the boundaries or beyond the parameter space covered by the \lacehc simulations (see Fig.~\ref{fig:dist_params}). While this situation did not pose significant issues for the fiducial emulator for $z>2$, it has a pronounced impact on the extended emulator, also at larger scales. Again, an avenue for enhancing these results would involve reducing the influence of scales with $k_{\parallel
}>4\ {\rm Mpc}^{-1}$. However, we defer this exploration to future investigations. 

The poor performance of the extended emulator in the \simigm highlights the need to expand the training sample parameter space by incorporating additional temperature and mean-flux rescalings that encompass universes with diverse IGM histories.  Another noteworthy observation is that the emulator predicts larger uncertainties for the snapshot at $z=2$ when compared to other more accurate predictions. This feature aligns well with the specific snapshot lying beyond the \lacehc parameter space.

\section{Conclusions}
\label{sec:conclusions}

In this paper, we build the first emulator of the one-dimensional \lya flux power spectrum (\poned) using a neural network (NN) architecture. To do so, we adopt the emulation strategy devised by \citetalias{Pedersen2021}, which relies on emulating \poned as a function of the amplitude and slope of the linear matter power spectrum on small scales. We summarise our main findings below:

\begin{itemize}

    \item In \S\ref{sec:emulator:nn}, we build an emulator that uses a Mixture Density Network (MDN) to predict the probability distribution of six polynomial coefficients describing \poned measurements. Then, it combines these distributions to generate predictions for the best-fitting solution and error prediction for each combination of input parameters. In Fig.~\ref{fig:net_characterisation}, we show how different decisions regarding the configuration of the emulator improve its precision. On the other hand, in Fig.~\ref{fig:sample_characterisation} we show that thanks to its MDN architecture, the emulator performance improves by 20\% when training it using \poned measurements from different simulation axes and phases instead of just relying on the average of these.

    \item In Fig.~\ref{fig:leave_one_out}, we show that the emulator precision is better than 1\% for cosmologies within the training set (leave-one-out tests) across the full range of scales ($0.1<k_\parallel[\iMpc]<4$) and redshifts ($2<z<4.5$) considered. Even though this value is similar to the precision quoted for the GP-based emulator described in \citetalias{Pedersen2021}, the actual performance of our emulator is better because we carry out a more detailed post-processing of the suite of hydrodynamical simulations from which these two studies extract \poned measurements.

    \item In Fig.~\ref{fig:other_cosmo}, we show that emulator predictions are largely insensitive to the impact of cosmic variance thanks to emulating on the space of polynomial coefficients \citep[see also][]{pedersen2023CompressingCosmologicalInformation}. Furthermore, we show that our emulator presents sub-percent precision for growth histories not included in the training set as well as for three $\Lambda$CDM extensions: massive neutrinos, running of the spectral index, and curvature. These findings confirm the advantage of emulating as a function of the amplitude and slope of the linear matter power spectrum on small scales rather than as a function of cosmological parameters.

    \item In Fig.~\ref{fig:P18k4}, we show that the emulator achieves on average 1.5\% precision for thermal and reionisation histories not considered in the training set. This is the consequence of emulating \poned as a function of the instantaneous properties of the IGM rather than parameters encoding information about the entire reionisation or thermal history of the universe.

    \item In Fig.~\ref{fig:extended_emulator}, we show the performance of an extended version of our emulator reaching $k_{\parallel}=8\iMpc$. In the \simcentral simulation, we find that the extended emulator presents an overall 1\% precision, a factor of $\sim$2 worse than the fiducial emulator. Its performance is especially bad for thermal and reionisation histories not considered in the training set, reaching on average 3.5\%.

    \item \lc{The \lya probe alone is not sufficient to establish competitive constraints on the extensions of the $\Lambda$CDM model. To effectively set meaningful constraints, we must combine \lya data with other external data sets, such as the Planck data \citep{pedersen2023CompressingCosmologicalInformation}. Our emulator is not designed solely for measuring cosmological parameters based on the \lyaf, rather to be integrated with other complementary probes, like the CMB, to derive more robust and accurate cosmological constraints.} 
    
\end{itemize}

As shown in \S\ref{sec:results}, the overall performance of the emulator is better for cosmologies and $\Lambda$CDM extensions not included in the training set than for reionisation histories not considered, which emphasises the importance of running simulations adopting distinct reionisation histories. This issue can be ameliorated by carrying out mean flux and temperature rescalings in a post-processing stage \citep{hydro_Lukic2015}, leading to a significant increase in the size of the training set. Such an increase makes NN-based models more suited to emulate \poned measurements than GP-based models because the runtime of the first and second increase linearly and with the cube of the number of input points, respectively.  \lc{For example, on a single CPU, both \texttt{LaCE-GP} and \texttt{LaCE-NN} take around one second to train on the 330 training points from \citep{Pedersen2021}. However, including the five mean-flux rescalings presented in this paper (\S\ref{sec:rescalings}),  \texttt{LaCE-GP} triples the training time of \texttt{LaCE-NN} (27s and 9s respectively). Extending the training sample using different combinations of axes and phases (\S\ref{sec:trainingsample_characterisation}), the training times for 2700 training points correspond to 147s for \texttt{LaCE-GP} and 20s for \texttt{LaCE-NN}. Nevertheless, it is worth noting that Gaussian Processes exhibit strong performance and fast training times on small data sets. Hence, we recommend considering their application before opting for a neural network in small data sets.}

We designed the fiducial version of our emulator aiming to analyse medium-resolution spectra from the DESI survey, which explains the range of redshifts and scales considered. In \S\ref{sec:extended_emulator}, we present an extended emulator conceived for the joint analysis of DESI and high-resolution measurements \citep[e.g.,][]{Karacayli2022} that reaches a factor of two smaller scales than the fiducial one. \lc{For this extended emulator, we find that the NN-based emulator performs significantly better than the GP-based emulator on small scales}, further highlighting the advantages of NNs for \poned emulation. On the other hand, we found the performance of the extended emulator is on average a factor of two times worse than the one of the fiducial emulator. In future work, we will further work on improving the precision of the extended emulator.

\lc{In this publication, we have not presented cosmological inferences using the emulator. When doing this, it will be important to discuss the fact that the emulator uncertainty varies as a function of emulator parameters.
This might have a non-negligible impact on the posteriors.
We defer this study to future work.}

Throughout this paper, we have focused on the one-dimensional \lya flux power spectrum. However, this statistic only contains part of the cosmological and astrophysical information encoded in the \lyaf as it neglects correlations between different line-of-sights. In future work, we will use the \lacehc simulations and an NN-based architecture to develop the first emulators for the \lya flux probability distribution \citep{lee2015IGMConstraintsSDSSIII}, one-dimensional bispectrum \citep{b1d_viel2009}, and three-dimensional power spectrum \citep{p3d_FontRibera2018}. A coherent analysis of \poned and these complementary statistics would enable fully exploiting the constraining power of the \lyaf.

\section*{Acknowledgements}
LCG, JCM and AFR acknowledge support from the European Union’s Horizon Europe research and innovation programme (COSMO-LYA, grant agreement 101044612).  AFR acknowledges financial support from the Spanish Ministry of Science and Innovation under the Ramon y Cajal program (RYC-2018-025210) and the PGC2021-123012NB-C41 project, and from the European Union's Horizon Europe research and innovation programme (COSMO-LYA, grant agreement 101044612). IFAE is partially funded by the CERCA program of the Generalitat de Catalunya. The analysis in this article have been performed at Port d'Informació Científica (PIC). We would like to acknowledge the support provided by PIC in granting us access to their computing resources.  The simulations were run using the Cambridge Service for Data Driven Discovery
(CSD3), part of which is operated by the University of Cambridge Research
Computing on behalf of the STFC DiRAC HPC Facility (\url{www.dirac.ac.uk}).
The DiRAC component of CSD3 was funded by BEIS capital funding via STFC capital
grants ST/P002307/1 and ST/R002452/1 and STFC operations grant ST/R00689X/1.
DiRAC is part of the National e-Infrastructure

\section*{Data Availability}
The simulations utilized for the development and testing of this work are publicly accessible via the following link: \url{https://github.com/igmhub/LaCE}.



\bibliographystyle{mnras}
\bibliography{emulatorP1d} 

\appendix
\section{Polynomial fit}
\label{sec:poly_order}

\begin{figure}
\centering
\includegraphics[width= 0.48\textwidth]{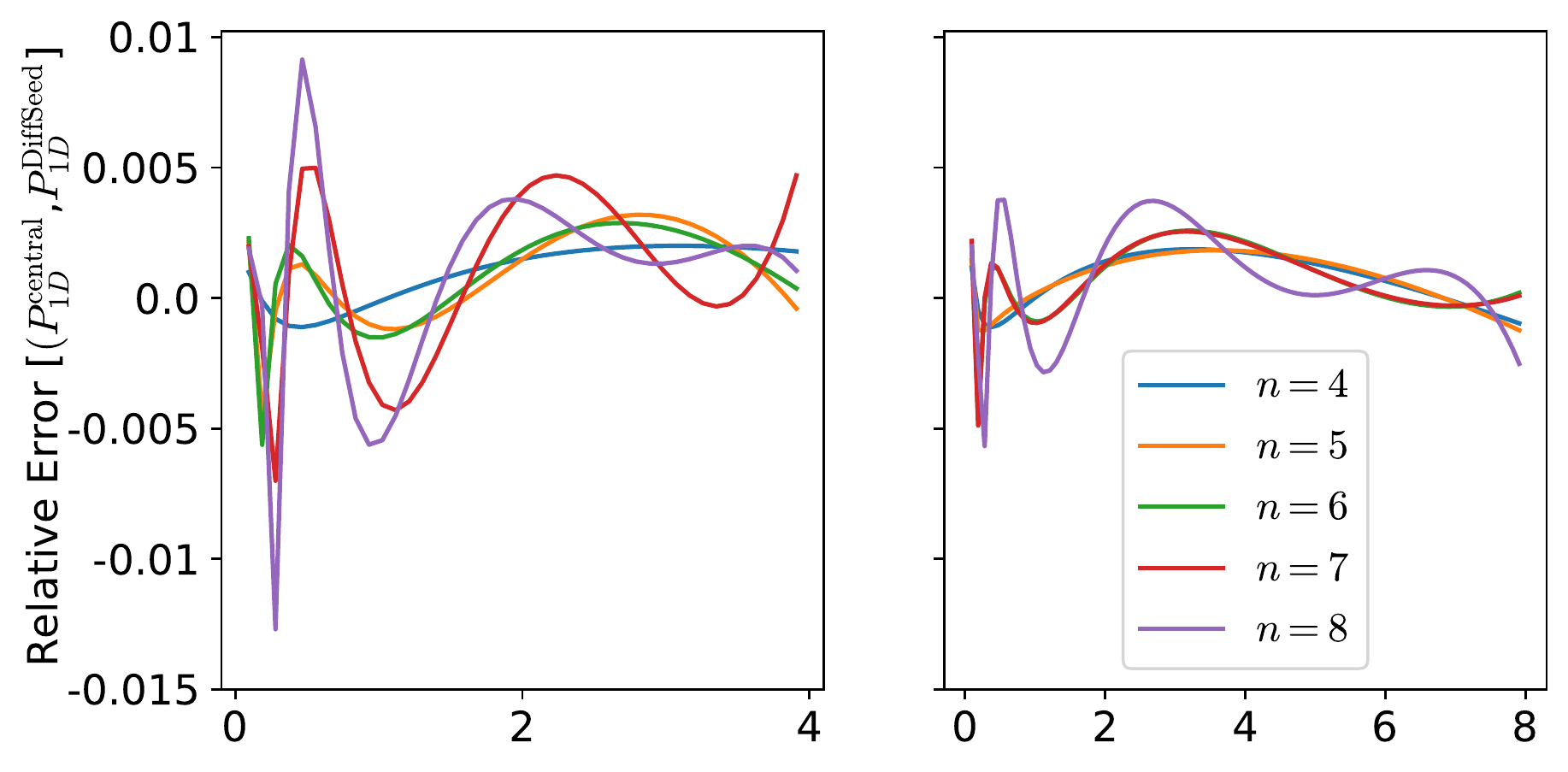}
\includegraphics[width= 0.48\textwidth]{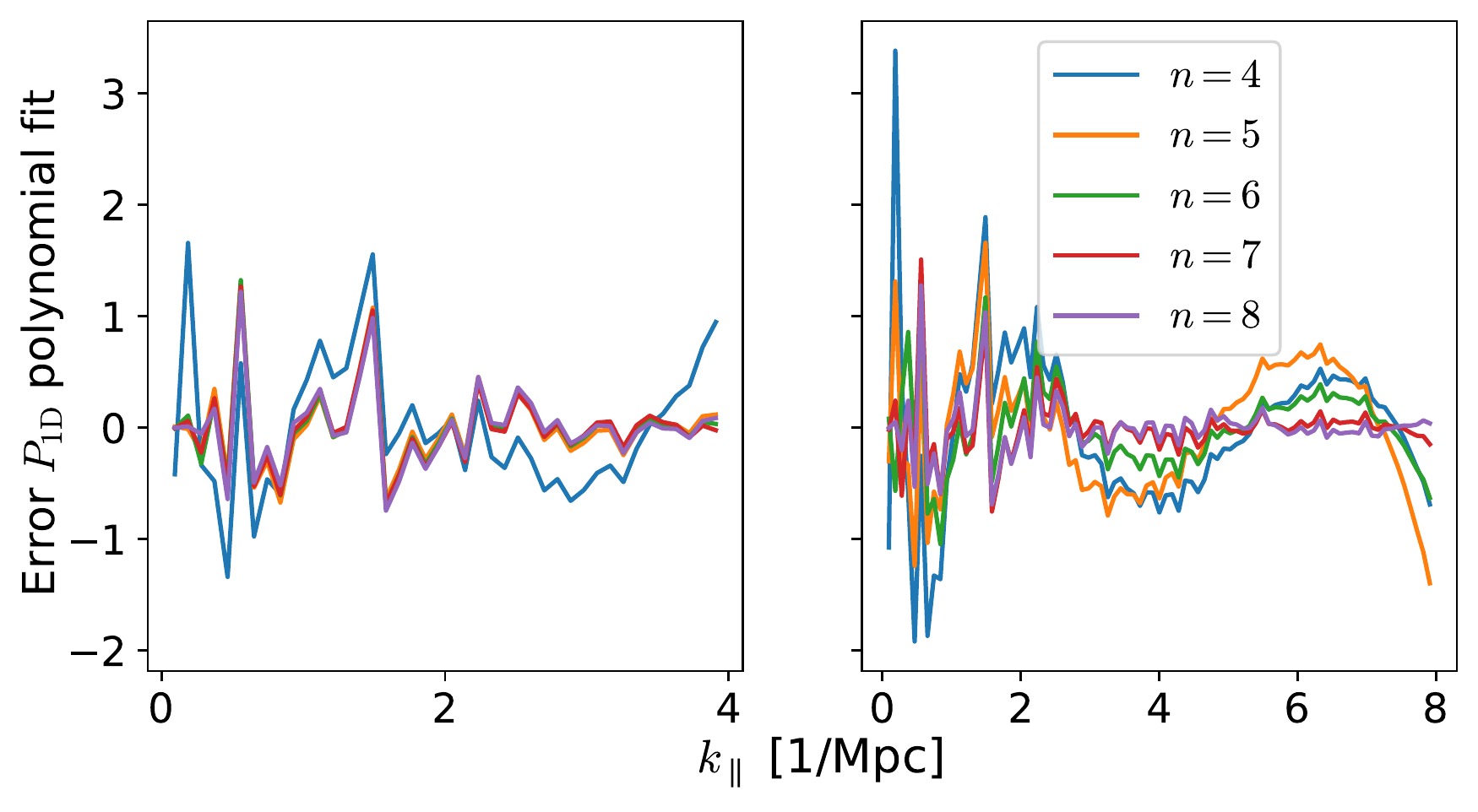}
\caption{Polynomial fit analysis to select the optimal polynomial order best fitting the \poned. The \emph{left} panels corresponds to fits to $k_{\parallel
}>4 \ {\rm Mpc}^{-1}$, while the \emph{right} ones expand to $k_{\parallel
}>8 \ {\rm Mpc}^{-1}$. The \emph{top} panels show the relative difference between the polynomial fits to the \poned of the central and the \simseed simulations. The \emph{bottom} panels show the percent error in the \poned polynomial fits compared to the true \poned.}
\label{fig:polyfit}
\end{figure}

As explained throughout the paper, \poned measurements in the training and test samples are smoothed using a polynomial fit and the emulator predicts the polynomial coefficients best fitting the smoothed \poned. Therefore, this type of emulator require choosing the adequate polynomial order. If a polynomial of lower order than required is used, it will oversmooth the \poned resulting in the loss of some important features. In contrast, a polynomial with a higher order than needed will overfit the data. For the fiducial emulator to $k_{\parallel
}<4 \ {\rm Mpc}^{-1}$, the polynomial is set to order $n = 5$. For the extended emulator (\S\ref{sec:extended_emulator}), where the polynomial must fit the \poned to $k_{\parallel
}<8 \ {\rm Mpc}^{-1}$, our choice is $n = 7$. In this appendix, we aim to justify this choices.

To select the appropriate polynomial coefficients, we conducted two tests. First, we fit the central and the \simseed test simulations (\S\ref{sec:methods_sims}) with polynomials of different orders. These two simulations have the same cosmological and IGM parameters, but a different set of Fourier phases for the initial conditions. Consequently, the \poned polynomial fits to both simulations should be identical, up to cosmic variance. The top plots in Fig.~\ref{fig:polyfit} shows the \lc{percent} error between the polynomial fits to the central and the \simseed simulations for polynomial orders within $n \in [4,8]$. The top left panel fits scales up to $k_{\parallel
}=4 \ {\rm Mpc}^{-1}$, while the right panels expands to $k_{\parallel
}=8 \ {\rm Mpc}^{-1}$. 

For the former case, $n=7$ and $n=8$ show larger fluctuations, which indicate that the polynomial is overfitting the data trying to fit cosmic variance. However, note that small fluctuations between the central and the \simseed simulations is necessary, but not sufficient to justify the polynimial order. A too low order polynomial could fit equally bad both simulations and give a very small \lc{percent} error. To select among the remaining possible orders, we check on the bottom left panel, which indicates the residuals in the \poned fit compared to the true \poned. In this case, we see that $n=4$ provides a larger error than the rest of the polynomial orders. Consequently $n=5$ is the lowest polynomial order satisfying both minimising the difference between the central and the \simseed simulation fits and the percent error compared to the true \poned. 

In the case of the extended emulator (right plots), it starts overfitting at $n=8$. The bottom plot indicates that $n=7$ has smaller percent errors than lower polynomial order, thus this is the selected polynomial order for the extended emulator.

\section{Covariance of the emulated \poned}
\label{sec:covariance}

\begin{figure}
    \centering
\includegraphics[width= 0.48\textwidth]{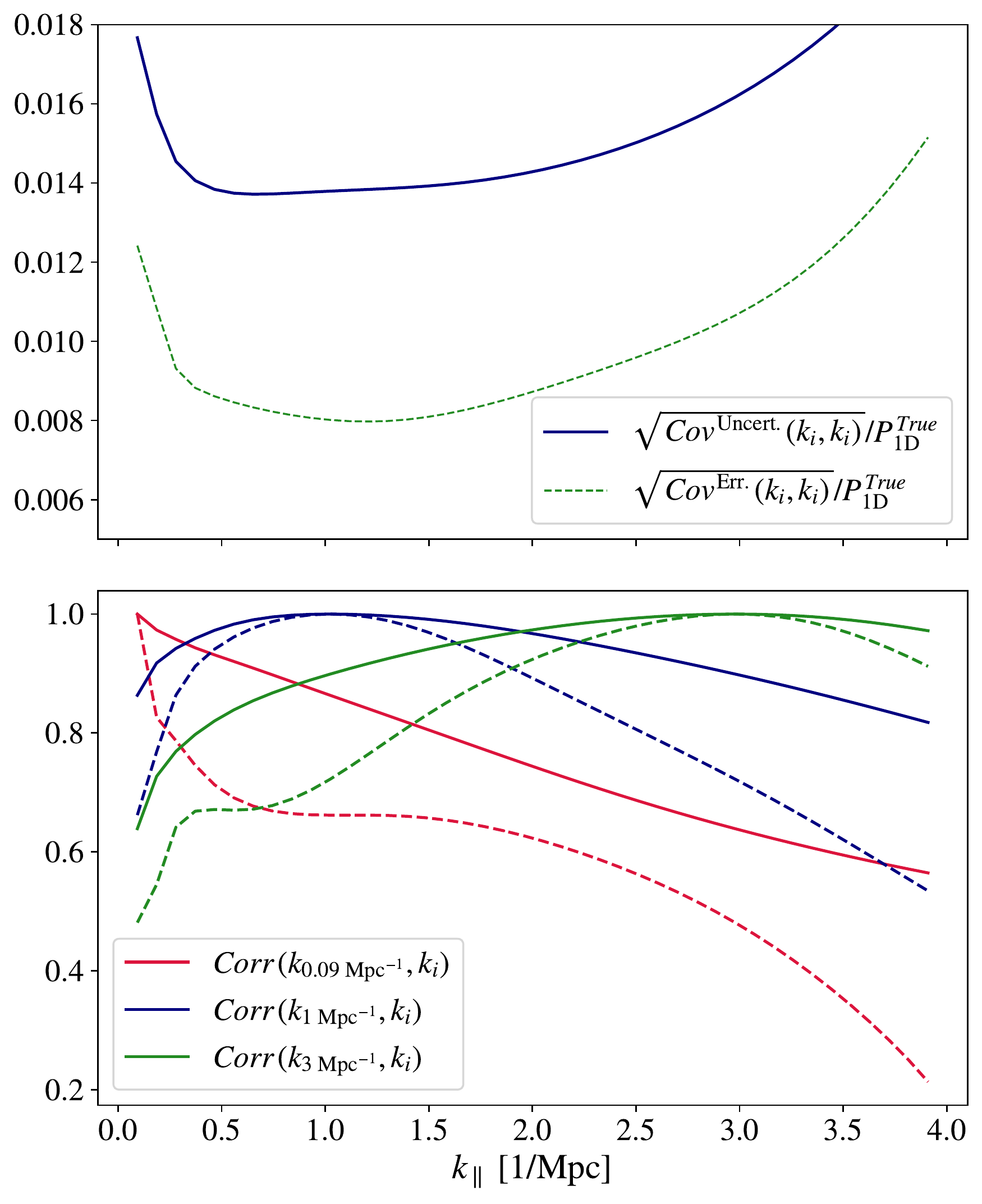}
\caption{Study of the \poned covariance obtained using MC sampling from the probability distribution of the polynomial terms. The \emph{top} panel shows the diagonal terms of $Cov^{\rm Uncert.}$ and $Cov^{\rm Err.}$ (blue and red lines from Eqs.~\ref{eq:cov_uncert} and \ref{eq:cov_err}. The \emph{bottom} panel shows the correlation coefficients fixing the scale in one direction and observing the variation along the other. The solid line depicts $Cov^{\rm Uncert.}$ and the dashed one $Cov^{\rm Err.}$. }
\label{fig:cov}
\end{figure}

The \texttt{LaCE-NN} employs a predictive approach by estimating the probability distribution of polynomial coefficients that best capture the target \poned. This distribution is subsequently utilised to derive an estimation of the \poned as well as its associated uncertainty (Eqs.~\ref{eq:p1d_polynomial} and \ref{eq:p1derr_polynomial}). These distributions also enable the estimation of covariance for the \poned measurements, providing insights into the correlations among different scales. In this study, we have undertaken an exploration of covariance measurement using Monte Carlo sampling \citep[MC,][]{MC_Robert2009}.

To generate the covariance matrix through MC sampling, we adopt the following procedure. For every \poned measurement, we generate 1000 realisations of the probability distribution for the polynomial coefficients, as estimated by the emulator. Subsequently, these realisations are employed to compute 1000 distinct \poned values for each $k_{\parallel}$ bin. These \poned values are then utilised to calculate the covariance measurement for each $k_{\parallel}$ bin.

In our current approach, we make the assumption that the polynomial coefficients are independent of each other. While we have trained the neural network by treating these coefficients as independent measurements in the loss function (\S\ref{sec:emulator:nn}), we recognise that this assumption is not entirely accurate. As a result, we acknowledge the need for future work to extend the emulator and account for correlations among the various polynomial terms.

We define two covariance matrices
\begin{align}
    \mathrm{Cov}^{\rm Uncert.} &= \langle P_{\rm 1D}^{\rm MC} - \overline{P_{\rm 1D}^{\rm MC}} \rangle \cdot \langle \overline{P_{\rm 1D}^{\rm MC}} - P_{\rm 1D}^{\rm MC} \rangle 
    \label{eq:cov_uncert} \\
    \mathrm{Cov}^{\rm Err.} &= \langle \overline{P_{\rm 1D}^{\rm MLE}}  - P_{\rm 1D}^{\rm True} \rangle \cdot \langle  P_{\rm 1D}^{\rm True} - \overline{P_{\rm 1D}^{\rm MLE}}  \rangle 
    \label{eq:cov_err}
\end{align}
 where $P_{\rm 1D}^{\rm MC}$ corresponds to the 1000 MC realisations of the \poned obtained sampling from the predicted distribution of the polynomial terms and $P_{\rm 1D}^{\rm MLE}$ is the maximum likelihood estimate of the \poned.  To eliminate the redshift evolution of \poned and enable the combination of covariance from multiple simulations and snapshots, we estimate the covariance of $P_{\rm 1D}^{\rm Pred}$/$P_{\rm 1D}^{\rm True}$. In this approach, the top panel of Fig.~\ref{fig:cov} displays the diagonal errors of $Cov^{\rm Uncert.}$ (blue line). The emulator tends to overestimate the uncertainties in the \poned predictions, with an observed factor of two in the overestimation. 
 
In the bottom panel of Fig.~\ref{fig:cov}, the correlation coefficients are depicted by fixing the scale in one direction and observing the variation along the other direction. The solid lines represent the correlation matrix $Cov^{\rm Uncert.}$, while the dashed lines correspond to $Cov^{\rm Err.}$. Notably, we observe a qualitative level of agreement between the two correlation matrices. Furthermore, there are substantial correlations across different scales. Since the \poned are estimated from six polynomial terms, these correlations along the $k_{\parallel
}$ scales arise due to the interplay between these terms.

While further investigations are necessary to enhance our understanding of \poned measurement covariances, we are pleased to note that, to the best of our knowledge, this study represents the first instance in which a complete emulator covariance is provided for \poned emulators, rather than solely focusing on the diagonal elements. By extending our analysis to include the full covariance matrix, we offer a more comprehensive characterisation of the uncertainties and correlations associated with \poned measurements, paving the way for more accurate and informative cosmological analyses in the future.

\section{PCA emulator}
\label{sec:pcaemu}
\begin{figure}
\includegraphics[width= 0.48\textwidth]{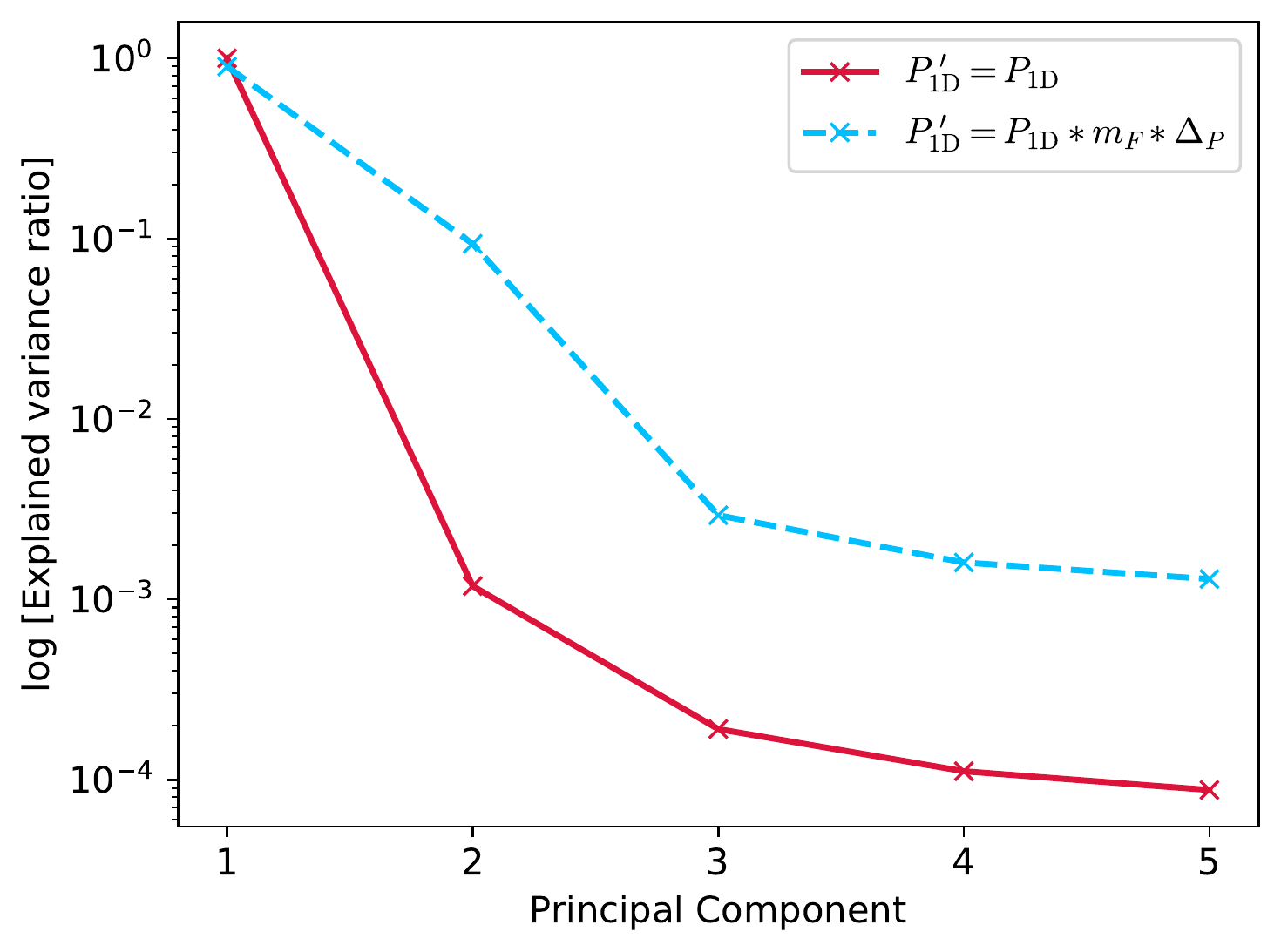}
\centering
\caption{Relative importance of the first five principal components obtained with a PCA decomposition of the \poned. The red line corresponds to the principal components of the \poned while the blue line shows the components of a scale version of the \poned aimed to reduce the amplitude among \poned (Eq.\ref{eq:pca_scaling}).}
\label{fig:PCA_variance}
\end{figure}
\begin{figure*}
\includegraphics[width= 0.98\textwidth]{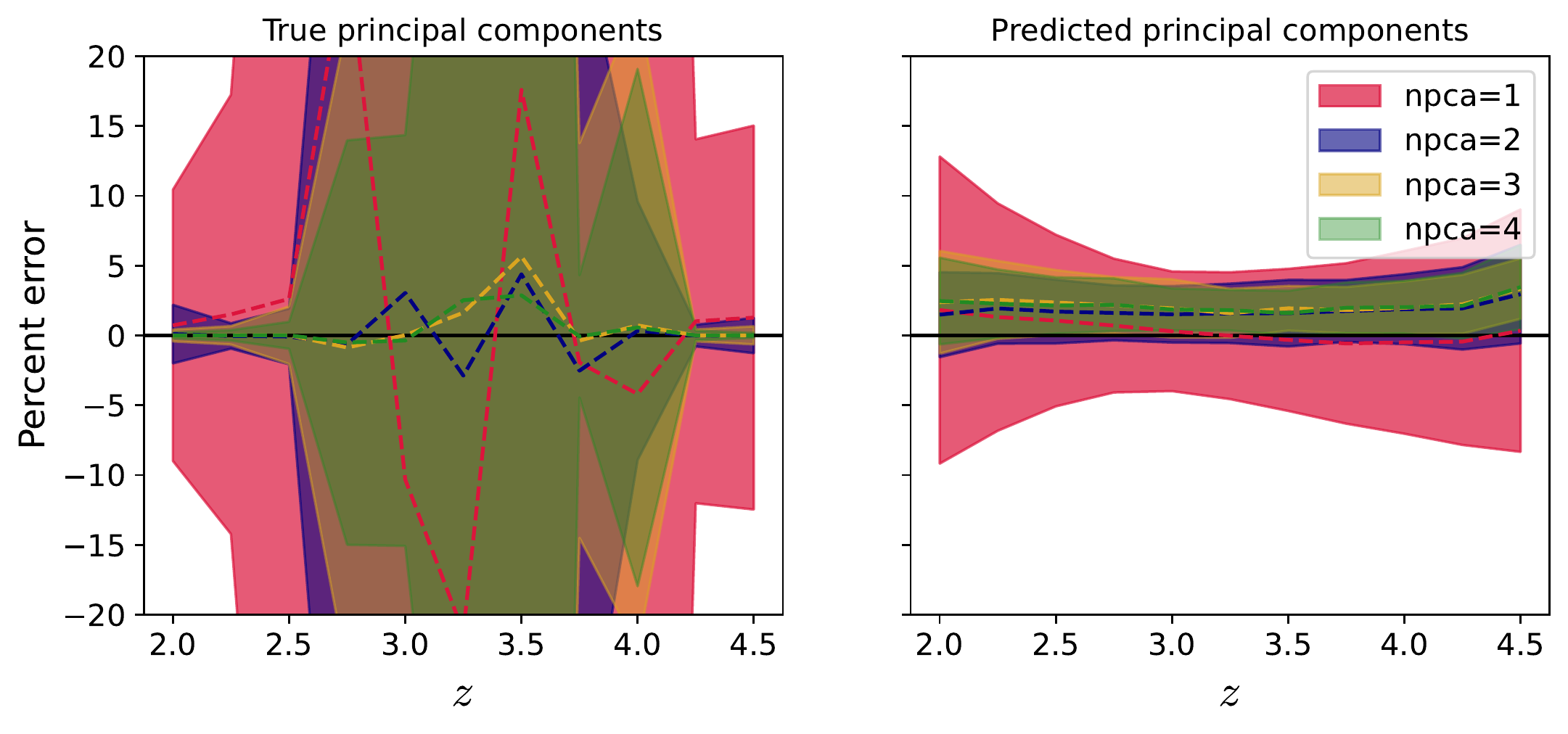}
\centering
\caption{Percent error in the \poned using PCA. Dashed lines correspond to the mean error while shaded areas indicate the standard deviation of the error predictions. On the \emph{left}, we show the results for the \poned reconstructed from the true principal components of a PCA decomposition. On the \emph{right}, the \poned are reconstructed from the emulated principal components  with the neural network PCA emulator. In all cases, the \poned are represented in the same parameter space.}
\label{fig:PCAemulator}
\end{figure*}

Principal Component Analysis \citep[PCA,][]{PCA_Pearson} is a widely used technique to extract the most significant patterns of variability from large data sets. In astrophysics, example PCA implementations include \citet{pcacosmo_ferreras} for studying the star formation history of elliptical galaxies in compact groups, and \citet{chaves-montero2020SurrogateModellingBaryonic, chaves-montero2021SurrogateModellingBaryonic} for the impact of the star formation history on galaxy colours. In cosmology, \citet{pcacosmo_campeti} studies how the shape of the spectrum of primordial gravitational waves can be constrained by future Cosmic Microwave Background (CMB) polarisation experiments and \citet{pcacosmo_sharma} uses PCA to reconstruct the dark energy equation of state from observational data of the Hubble parameter and Supernovae type Ia. In this section, we explore the possibility of PCA-based \poned emulator. 

This emulator consists of a neural network predicting the PCA components that best reconstruct the \poned under the transformation matrix converting from \poned space to PCA space. First, we decompose the training sample \poned using PCA and define the transformation matrix from $k$-space to principal component space. 
During the training phase, the emulator predicts the principal components, which are then propagated to a predicted \poned using such transformation matrix. Then, as in the polynomial case, the emulator is trained in \poned space using  Eq.~\ref{eq:loss_function}. This should enable the network to learn the importance that each principal component has on the \poned and weight the learning accordingly. 

Fig.~\ref{fig:PCA_variance} presents the relative importance of the first five principal components that represent the endpoints of the Latin hypercube (\S\ref{sec:methods_sims}). The red line denotes the contribution of each component, with the first principal component accounting for more than 99\% of the variance within the sample. While this suggests that a single principal component is sufficient to accurately represent and recover the \poned, our analysis reveals that most \poned require additional components for precise representation. Given the varying amplitudes of each \poned, we hypothesise that the first principal component encodes primarily the amplitude of the \poned.

In order to capture more features with the PCA decomposition, we rescale the \poned to minimise amplitude differences among them. For that, we fit 
\begin{equation}
    P^{\prime}_{\rm 1D} = P_{\rm 1D} \cdot \Delta_p \cdot m_{F}\,
    \label{eq:pca_scaling}
\end{equation} where $\Delta_p$ is the amplitude of the matter power spectrum and and $m_{F}$ is the neutral hydrogen mean flux (\S\ref{sec:methods_sims}). As for the polynomial fitting case, we fit the $\log_{10}$\poned in $\log_{10}k_{\parallel}$ (Eqs.\ref{eq:p1d_polynomial}).

In the blue dashed line in Fig.~\ref{fig:PCA_variance}, we show the relative importance of the principal components of $P^{\prime}_{\rm 1D}$. In this case, the relative importance of the first principal component is reduced, and the second component gains importance. Nevertheless, the first component still accounts for $\sim 90\%$ of the variance. 

Fig.\ref{fig:PCAemulator} shows the error in the \poned when these are constructed using principal components. Each colour corresponds to a \poned reconstruction with a different number of principal components, e.g. the red line corresponds to a single component while the green one, to four. The dashed lines indicate the mean error in the \poned, while the shaded area is the standard deviation of the error. 

The left panel presents the percent error of the \poned measured from the true principal components obtained directly from the PCA decomposition, prior to emulation. As shown in Fig,\ref{fig:PCA_variance}, a single principal component accounts for roughly 90\% of the variance in the sample. However, we observe that a single component is insufficient to recover the \poned accurately, resulting in a large error of $>20\%$. By utilising two, three, and four components we can reduce the error to approximately 5\%. Despite this improvement, it's worth noting that there remains a large scatter in the results, indicating that some \poned are not accurately represented by the principal components.

Despite the challenges in accurately recovering the \poned using the 'true' principal components, we have developed a PCA emulator for our data set. In the right panel of Fig.~\ref{fig:PCAemulator}, we present the error in the \poned obtained from the emulated principal components. We observe a significant reduction in error compared to the use of true principal components (left panel). This is possible because the emulator is trained in \poned space and it is not forced to recover the exact true PCAs, but minimising the \poned error. Nevertheless, note that we are recovering the \poned using the same transformation matrix as for the true principal components, meaning that we are emulating the \poned is the same parameter space. The fact that the emulated PCAs predict the \poned more accurately than the true ones suggests that the PCA emulator can capture higher-order correlations and additional information beyond what is possible with traditional PCA decomposition. While the PCA components themselves are a linear transformation of the original \poned data, the emulator introduces non-linearities into the prediction process, enabling greater flexibility in the prediction of principal components. Despite these improvements, the accuracy of the PCA emulator does not reach that achieved with the polynomial fit (\S\ref{sec:L1O}).

\section{Effect of the cosmological and astrophysical parameters on the \poned}
\begin{figure*}
\includegraphics[width= 0.98\textwidth]{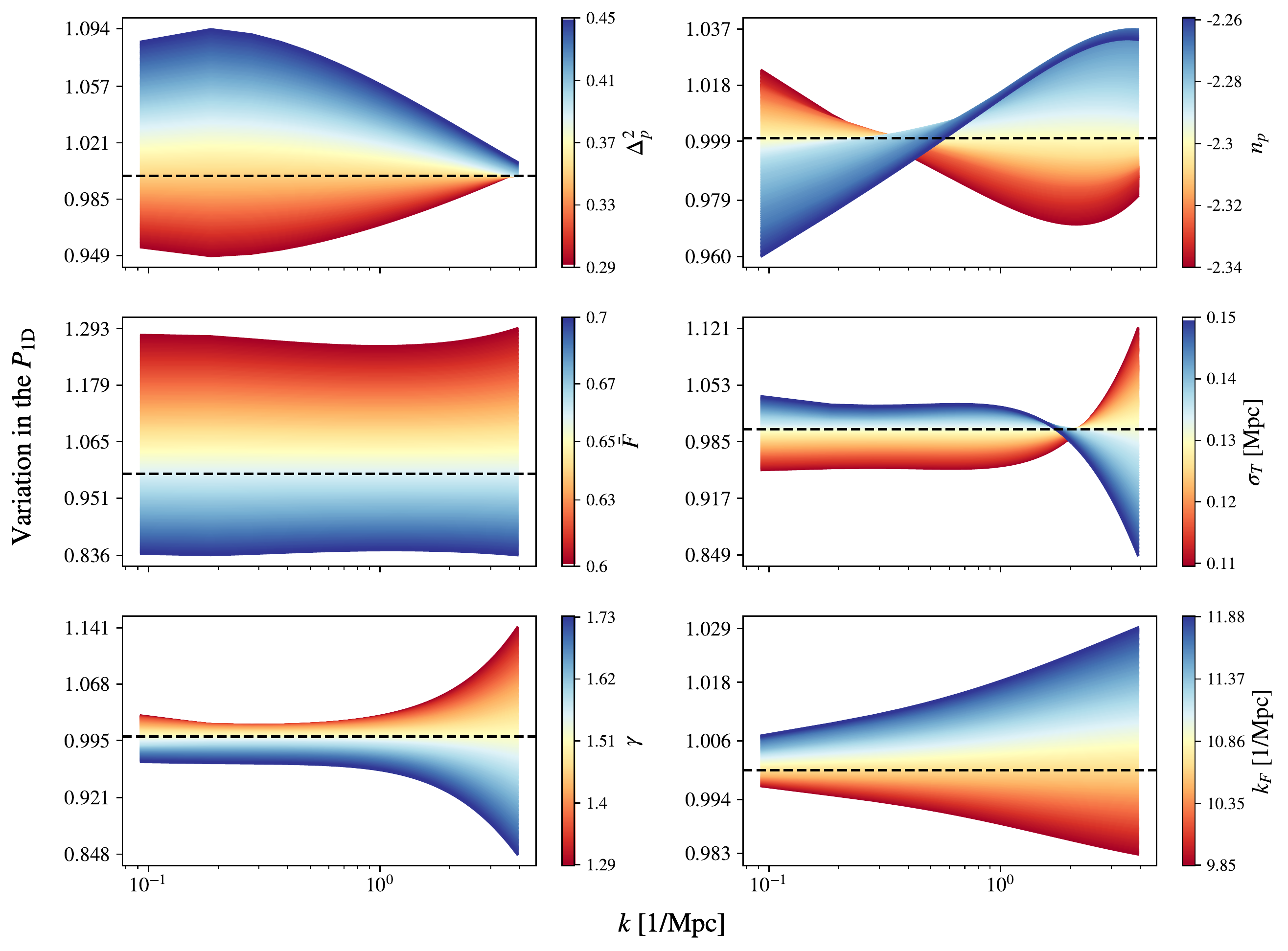}
\centering
\caption{Effect that changing the six input emulator parameters from edge to edge of the convex hull has on the \poned. We show this for the \simcentral at $z=3$. The colour scheme represents the lowest values for each parameter in red, and the highest in blue.}
\label{fig:smoothness_p1d}
\end{figure*}

\lc{This appendix investigates the emulator's response to changes in individual parameters. Figure \ref{fig:smoothness_p1d} illustrates the variation of the \poned for the central simulation at redshift $z=3$. In this plot, each simulation parameter is varied from edge to edge of the convex hull of the training set (see Fig. \ref{fig:dist_params}), while the other parameters remain fixed. This plot corresponds to the neural network version of Fig.~9 in \citet[][]{pedersen2023CompressingCosmologicalInformation}.

Similar to the \texttt{LaCE-GP} approach, our emulator exhibits a smooth response to variations in all emulator parameters. This smoothness is a crucial characteristic that ensures the robustness of the emulator's predictions across a wide range of parameter configurations. }

\bsp	
\label{lastpage}
\end{document}